\documentclass[12pt]{article}
\usepackage[T2A]{fontenc}
\usepackage[utf8]{inputenc}
\usepackage{amsthm, amsfonts,amssymb,amscd,cite}
\usepackage[english]{babel}
\usepackage[tbtags]{amsmath}
\usepackage[pdftex]{graphicx}
\usepackage{graphicx}
\usepackage{hyperref}

\let\cal\mathcal

\DeclareSymbolFont{bletters}{OML}{cmm}{bx}{it}
\DeclareMathSymbol{\blad}{\mathord}{bletters}{'025}
\DeclareMathSymbol{\bla}{\mathord}{bletters}{'013}
\DeclareMathSymbol{\bmu}{\mathord}{bletters}{'026}
\DeclareMathSymbol{\bnu}{\mathord}{bletters}{'027}
\DeclareMathSymbol{\bth}{\mathord}{bletters}{'022}
\DeclareMathSymbol{\bfI}{\mathord}{bletters}{"49}
\DeclareMathSymbol{\bdl}{\mathord}{bletters}{"0E}
\DeclareMathSymbol{\bDl}{\mathord}{bletters}{"001}
\def \bpi{\boldsymbol\pi}
\def \bphi{\boldsymbol\phi}

\begin{document}
\title{$${}$$\\
{\bf Heisenberg $XX$ chain,
non-homogeneously parameterised generating exponential, and diagonally restricted plane partitions}
}
\author{
{\bf\Large C.~Malyshev, N.~M.~Bogoliubov}\\
$${}$$\\
{\it Steklov Institute of Mathematics}
{\it (St.-Petersburg Department)}\\
{\it Fontanka 27, St.-Petersburg, 191023, RUSSIA}
}

\date{}

\maketitle

\def \bbe{\boldsymbol\be}
\def \bchi{\boldsymbol\chi}
\def \bdl{\boldsymbol\dl}
\def \Bdl{\boldsymbol\Dl}
\def \bphi{\boldsymbol\phi}
\def \bPsi{\boldsymbol\Psi}
\def \bsi{\boldsymbol\si}
\def \bta{\boldsymbol\eta}
\def \bvphi{\boldsymbol\varphi}
\def \bxi{\boldsymbol\xi}

\def \al{\alpha}
\def \be{\beta}
\def \ga{\gamma}
\def \dl{\delta}
\def \ep{\varepsilon}
\def \nb{\nabla}
\def \tet{\theta}
\def \ka{\varkappa}
\def \la{\lambda}
\def \si{\sigma}
\def \ph{\varphi}
\def \om{\omega}
\def \vt{\vartheta}
\def \z{\zeta}
\def \ta{\theta}

\def \Ga{\Gamma}
\def \Dl{\Delta}
\def \La{\Lambda}
\def \Si{\Sigma}
\def \Ph{\Phi}
\def \Om{\Omega}

\def \cA{\cal A}
\def \cB{\cal B}
\def \cC{\cal C}
\def \cD{\cal D}
\def \cE{\cal E}
\def \cG{\cal G}
\def \cN{\cal N}
\def \cI{\cal I}
\def \cT{\cal T}
\def \cR{\cal R}
\def \cF{\cal F}
\def \cY{\cal Y}
\def \cK{\cal K}
\def \cJ{\cal J}
\def \cX{\cal X}
\def \cZ{\cal Z}
\def \cQ{\cal Q}
\def \cW{\cal W}
\def \cM{{\cal M}}
\def \cL{{\cal L}}
\def \CU{{\cal U}}
\def \cS{{\cal S}}
\def \CK{\mathcal M}
\def \CM{\mathcal M}
\def \CN{\mathcal N}
\def \CV{\mathcal V}

\def \bA{\bold A}
\def \bB{\bold B}
\def \bC{\bold C}
\def \bD{\bold D}
\def \bE{\bold E}
\def \bP{\bold P}
\def \bR{\bold R}
\def \bU{\bold U}
\def \bV{\bold V}
\def \bH{\bold H}

\def\Ba{\rm{B}}
\def\Fa{\rm{F}}
\def \BI{\mathbb{I}}
\def \BC{\mathbb{C}}
\def \BD{\mathbb{D}}
\def \BZ{\mathbb{Z}}
\def \BR{\mathbb{R}}
\def \BQ{\mathbb{Q}}
\def \BN{\mathbb{N}}
\def \IM{\Im}
\def \RE{\Re}
\def \1{^{-1}}
\def \cd{\partial}
\def \at{{\rm arctan}\,}
\def \ch{{\rm ch}\,}
\def \sh{{\rm sh}\,}
\def \th{{\rm th}\,}
\def \bg{{\rm bg}\,}
\def \e{{\rm e}\,}
\def \c{{\rm c}\,}
\def \m{{\rm m}\,}
\def \dr{{\rm d}\,}
\def \o{{\rm o}\,}
\def \ld{\ldots}

\def \w{\widetilde}
\def \h{\widehat}
\def \t{\times}
\def \l{\langle}
\def \r{\rangle}
\def \Tr{{\rm Tr}\,}
\def \tr{{\rm tr}\,}
\def \diag{{\rm diag}\,}
\def \row{{\rm row}\,}
\def \Det{{\rm det}\,}
\def \Ddet{\text{\rm Det}\,}
\def \d{\dagger}
\def \pprime{\prime \prime}
\def \babe{\bar\beta}
\def \nt{{\widetilde n}}

\begin{abstract}
\noindent  The mean values of non-homogeneously parameterized generating exponential are obtained and investigated for the periodic Heisenberg $XX$ model. The norm-trace generating function of boxed plane partitions with fixed volume of their diagonal parts is obtained as $N$-particles average of the generating exponential. The
generating function of self-avoiding walks
of random turns vicious walkers is obtained in terms of the circulant matrices that leads to generalizations of the Ramus's identity. Under various specifications of the generating exponential, the $N$-particles averages arise for a set of inconsecutive flipped spins and for powers of the first moment of flipped spins distribution at large length of the chain. These averages
are expressed through the numbers of closed trajectories with constrained initial/final positions. The estimates at large temporal parameter are expressed through the numbers of diagonally restricted plane partitions characterized by fixed values of the main diagonal trace or by fixed heights of the diagonal columns in one-to-one correspondence with the flipped spins positions.
\end{abstract}

\vskip1.0cm
\noindent{\emph{{\bf Keywords:}} } symmetric functions, plane partitions, self-avoiding lattice walks, circulant matrix, Ramus's identity

\thispagestyle{empty}
\newpage

\section{Introduction}
\label{sec1}

Mathematical methods developed in quantum integrable models \cite{bax, fad} find application in different branches of physics
\cite{Nekr, nakat, forrmill, Shor, resh, boro, foda, kitpr, zinnu, essl, goh, rzz}. The Quantum Inverse Scattering Method \cite{fad, KBI2} provides a powerful approach to calculation of the correlation functions of the one-dimensional spin-$1/2$ anisotropic $XXZ$ model \cite{vk, vk1, slav, ml1, kit1, ml22}.

The $XX$ chain is the free-fermion limit of the $XXZ$ model. Despite its simplicity, the model is attractive from different perspectives.
In fact, it provides a base for studying of entanglement entropy as a measure of entanglement \cite{Sugino}. Connection between the $XX$ chain and the low-energy QCD, as well as a possibility of a third order phase
transition \cite{gross} in the spin chain, are discussed in \cite{tier, zah}.
Intriguing relationship of the model in question with the integrable combinatorics \cite{bor, bmumn} attracts special attention.
The temperature correlation functions in the $XX$ chain were calculated  and studied in the thermodynamical limit in \cite{col, col1,m1}.

Our approach to the investigation of correlation functions is based on the theory of symmetric functions \cite{macd}, which allows us to establish natural connection with the different types of the directed lattice walks, partitions and plane partitions \cite{kratt, bres}. In \cite{b1, b11} it was shown that the multi-spin correlation functions over the ferromagnetic vacuum are in one-to-one correspondence with the path configuration of the random turns walkers \cite{fish, forr2}. The correlation functions calculated over the ground state lead to the more complicated structure of the lattice paths \cite{bmumn, bmnph, b2, b3}.

The enumeration of plane partitions with the different constrains is a classical part of the enumerative combinatorics \cite{stan1},
and their number with the fixed values of diagonal parts is of particular interest \cite{st}. The temporal evolution of the
first moment of particles distribution of the phase model \cite{bojpa} after special $q$-parametrization coincides with the norm-trace generating function \cite{statm}, while the partition function of the four vertex model in the linearly growing external field  under the so called ``scalar product'' boundary conditions
counts plane partitions with the fixed values of their diagonal parts \cite{bmjpa}.

In the present paper we shall consider the \textit{generating exponential
operator} $\exp{\cQ}$, where $\cQ =\frac12\sum_{k=1}^M \al_k (1- \si^z_k)$ is the weighted inhomogeneous sum of flipped spins with the parameters
$\al_k$ depending on the lattice sites. The average of $\exp{\cQ}$ over $N$-particles ground state is represented in the determinantal form. The obtained answer allows to derive the
generating function of boxed plane partitions with the fixed sums of their
diagonals. The generating function of $N$ random turns walkers is expressed
in terms of the entries of products of the circulant matrix \cite{dav, cray}. The Ramus's identity \cite{ram} and its multiple series
generalizations enable to obtain identities respected by the numbers of
$K$-steps lattice paths of $N$ vicious walkers. These identities are used then to obtain the temporal correlation functions of inconsecutive flipped
spins in terms of the superposition of the nests of self-avoiding lattice
paths.

\textit{Organization of the paper}. Section~\ref{sec1} is introductory. The outline is given by Section~\ref{sec2}.
The $N$-particles Bethe state-vectors expressed through the Schur functions and a combinatorial interpretation of the Schur functions in terms of nests of self-avoiding lattice paths are presented in Section~\ref{sec5}. The norm-trace generating function of the boxed plane partitions with fixed sums of their diagonal parts is derived in Section~\ref{sec53}.
Section~\ref{50} is devoted to the transition amplitude over $N$-particles states which respects the differential-difference equation.
Solution to a descendant difference equation is obtained
in terms of the circulant matrices expressed through the lacunary sums of the binomial coefficients. The power series representation for the generating function of random turns walks of vicious walkers is obtained.
The multiple series generalizations of Ramus's identity are derived in Section~\ref{50}.
The Boltzmann weighted average of the generating exponential and its relationship with the lattice walks are considered in
Section~\ref{ss342}.
The temporal correlation functions of flipped spins are obtained in Section~\ref{sec60} and their combinatorial interpretation is given in terms of enumeration of self-avoiding lattice walks and of diagonally restricted plane partitions. The $N$-particles   Boltzmann-weighted mean values are obtained for the generating exponential, for a projector onto a set of inconsecutive flipped spins, and for powers of the first moment of flipped spins distribution at large enough length of the periodic chain. The estimates at large temporal parameter are obtained in terms of enumeration of boxed plane partitions with diagonal elements subjected to additional restrictions. Discussion in Section~\ref{sec6} completes the paper.

\section{Outline of the problem}
\label{sec2}

The $XX$ Heisenberg spin chain is described by the Hamiltonian:
\begin{eqnarray}
H=H_{\rm xx} -h S^z\,,\qquad H_{\rm xx}\equiv - \,\frac12 \sum \limits^{M}_{n, m=1} \Dl_{nm}\si^+_{n}\si^-_{m}\,, \label{cor:lin2}
\\
S^z=\frac12\sum \limits_{n=1}^M \si^z_n\,, \label{cor:lin3}
\end{eqnarray}
where $S^z$ is the third component of total spin, $h\ge 0$ is homogeneous magnetic field, and the number of sites is $M=0\pmod{2}$. The local spin operators
$\si^\pm_n=\frac12 (\si^x_n \pm i\si^y_n)$ and $\si^z_n$ depend on the lattice argument
$n\in\cE \equiv\{1, 2, \dots, M\}$, act on the state space  $\frak{H}_M \equiv ({\BC}^2)^{\otimes M}$,
and satisfy the commutation relations:
\begin{equation}
[\si^+_k, \si^-_l]\,=\,\dl_{k l}\,\si^z_l\,,\qquad
[\si^z_k,\si^\pm_l]\,=\,\pm2 \dl_{k l}\,\si^{\pm}_l\,.
\label{comm}
\end{equation}
The entries $\Dl_{n m}$ (\ref{cor:lin2}) constitute $M\times M$ \textit{hopping matrix} $\bold\Delta$ and are of the form:
\begin{equation}
\Dl_{nm}\,\equiv\,
\dl_{|n-m|, 1} + \dl_{|n-m|, M-1}\,,
\label{cor:lin4}
\end{equation}
where ${\dl}_{n, l} (\equiv {\dl}_{n l})$ is the Kronecker symbol. The matrix ${\bold\Delta}$ is a special type of so-called \textit{circulant} matrix \cite{dav, cray}. The periodic boundary conditions
$\si^{\#}_{n+M}=\si^{\#}_n$, $\#\in\{\pm, z\}$, $\forall n\in {\cE}$, are imposed, and the Hamiltonian $H$ (\ref{cor:lin2}) commutes with $S^z$.

Spin ``up'' and ``down'' states on $n^{\rm{th}}$ site, $\mid \uparrow \rangle_n$
and $\mid \downarrow \rangle_n$,
are defined so that the rising/lowering operators $\si_n^{\pm}$ act on them as follows:
\begin{equation}
\label{cor:lin411}
\si_n^{+} \mid \downarrow \rangle_n\,=\,\mid \uparrow \rangle_n\,,
\qquad
\si_n^{-} \mid \uparrow \rangle_n\,=\,\mid \downarrow \rangle_n\,,\qquad
\si_n^{-} \mid \downarrow \rangle_n = \si_n^{+}\mid \uparrow \rangle_n =0\,.
\end{equation}
From (\ref{cor:lin411}) it follows that two operators ${\sf q}_n$ and ${\bar{\sf q}}_n$,
\begin{equation}
\label{cor:lin412}
{\sf q}_n \equiv \si_n^{-} \si_n^{+} = \frac12 (1- \si^z_n)\,,\quad {\bar{\sf q}}_n \equiv \si_n^{+} \si_n^{-} = \frac12 (1 + \si^z_n)\,,
\end{equation}
are the local projectors since ensure
\begin{equation}
\label{cor:lin413}
{\sf q}_n \mid \downarrow \rangle_n = \mid \downarrow \rangle_n\,,\quad
{\sf q}_n \mid \uparrow \rangle_n = 0\,,\qquad {\bar{\sf q}}_n
\mid \uparrow \rangle_n = \mid \uparrow \rangle_n\,,\quad
{\bar{\sf q}}_n
\mid \downarrow \rangle_n = 0\,.
\end{equation}

The state $\mid\Uparrow\rangle
\equiv \bigotimes_{n=1}^{M} \mid \uparrow \rangle_n$ (spins ``up'' on all sites) is chosen as the reference state (i.e., pseudovacuum \cite{KBI2}),
and therefore the reversed spin on $n^{\rm{th}}$ site $\mid \downarrow \rangle_n$ will be called \textit{flipped} spin. Regarding (\ref{cor:lin413}), the sum $Q(m)\equiv \sum_{k=1}^{m}
{{\sf q}}_k$ is the number of flipped spins operator on first $m$ sites. The total number of flipped spins operator is $\cal N \equiv Q(M)$, and it commutes with $H$ (\ref{cor:lin2}).

Let us introduce the sum of ${\sf q}_n$ (\ref{cor:lin412}) taken with the ``weights'' $\al_n\in\BC$,
\begin{equation}
\cQ\equiv \sum_{n=1}^M
\al_n {\sf q}_n\,,
\label{cor:lin05}
\end{equation}
and let us consider the mean value of the generating exponential
operator $e^{\cQ}$:
\begin{equation}
{{\langle\langle}} e^{\cQ} {{\rangle\rangle}}_{\be}\,
\equiv\, {\sf trace}\,(e^{\cQ}\boldsymbol{\rho} )\,,\qquad \boldsymbol{\rho} \equiv \frac{e^{-\be H}}{{\sf trace}\,(e^{-\be H})}
\,,
\label{cor:lin5}
\end{equation}
where $\be$ is a real positive parameter, the Hamiltonian $H$
is given by (\ref{cor:lin2}), (\ref{cor:lin3}),
and $\boldsymbol{\rho}$ is density matrix. The parameter $\be$ might be treated either as an ``evolution'' parameter \cite{b1, statm} or inverse absolute temperature. The trace symbol in (\ref{cor:lin5}) implies summation over states of the model and will be concretized in Section~\ref{sec60}.

Generating functions provide a helpful tool for derivation of certain correlation functions of the quantum integrable models \cite{KBI2}. The operator $e^{\cQ}$ is called `generating exponential' since
$\langle\l e^{\cQ} \r\rangle_{\be}$ (\ref{cor:lin5}) parameterized by the elements of $M$-tuple
${\bf a}_M \equiv (\al_{1},
\al_{2}, \ldots ,
\al_{{M}})$ can be viewed as the
generating function $G ({\bf a}_M) \equiv \l\langle e^{\cQ} \r\rangle_{\be}$ of the
mean values of products ${\varPi}_{\bf k} \equiv \prod_{j=1}^{l} {{\sf q}}_{k_j}$ of the flipped spins projectors ${\sf q}_n$ (\ref{cor:lin412}):
\begin{align}
\l\l {\varPi}_{\bf k} \r\r_{\be} &=\,{\sf trace}\,\bigl({\varPi}_{\bf k}\, \boldsymbol{\rho}\bigr)
\label{cor:lin611}\\
&=\, \lim\limits_{\{\al_k \to 0\}}
\frac{{\cd}^l\, G ({\bf a}_M)}{\cd \al_{k_1} \cd \al_{k_2}\,\dots\,\cd \al_{k_l}}
\equiv \lim\limits_{\{\al_k \to 0\}} {\cd}^l_{\al_{k_1} \al_{k_2} \ldots \al_{k_l}} G ({\bf a}_M)\,,
\label{cor:lin61}
\end{align}
where $1 \le k_1 < k_2 < \dots < k_l \le M$. The product ${\varPi}_{\bf k}$ is the projector onto $l$ inconsequent flipped spins. Recall that the correlation functions of string $\bar{\varPi}_l \equiv \prod_{j=1}^{l} {\bar{\sf q}}_j$
of the projectors ${\bar{\sf q}}_j$ (\ref{cor:lin412})
and appropriate combinatorial implications have been studied in \cite{bmumn, nest, b3}.

When the elements of ${\bf a}_M$ depend linearly on the site coordinate,
${\bf a}_M = \al \cdot ({1},
{2}, \ldots , {{M}})$, $\al\in\BR$, the operator ${\cal Q}$ is reduced to ${\cal Q} = \al{\sf M}$, where ${\sf M}$ would be considered as the first moment of flipped spins distribution. The mean values of $e^{\al{\sf M}}$ are the generating functions of the mean values of powers of ${\sf M}$:
\begin{equation}\label{new30}
\l\l {\sf M}^l \r\r_{\be} \,=\,
\lim\limits_{\{\al \to 0\}} {\cd}^l_{\al} \l\l e^{\al{\sf M}} \r\r_{\be}\,,\qquad {\sf M} \equiv
\sum_{n=1}^{M} n\,{\sf q}_n\,.
\end{equation}
The temporal evolution of $e^{\cQ}$ for ${\cQ}$ proportional to the first moment of particles distribution has been studied in \cite{statm} for the quantum phase model.

With the aim of evaluation of $\l\l e^{\cQ} \r\r_{\be}$ (\ref{cor:lin5}), $\l\l {\varPi}_{\bf k} \r\r_{\be}$ (\ref{cor:lin611}), and $\l\l {\sf M}^l \r\r_{\be}$ (\ref{new30}), the approach based on symmetric functions \cite{bmnph, bmumn}
is developed in the present paper
to derive ${\sf trace}\,(e^{\cQ} e^{-\be H} )$ and relate it, at large enough $M$, with enumeration of random turns walks of vicious walkers occupying specially prescribed initial/final positions on the chain and with enumeration of boxed plane partitions subjected to certain restrictions.

When $\al_1 = \al_2 =\dots = \al_m = \al$ and
$\al_{m+1} = \al_{m+2} =\dots = \al_M = 0$ (conventional choice), the operator ${\cal Q}$ takes the form ${\cal Q}=\al Q(m)$, and $G({\bf a}_M)$ coincides with the generating function $G(\al, m)$ of the correlation functions of $z$-components of spins for the Heisenberg chains \cite{vk1, kit1, col1, ess}. The `emptiness formation probability', being probability of formation of (ferromagnetic) string of $m$ consecutive ``up'' spins is given by $G(\al, m)$ at $\al\to -\infty$, \cite{ess, ml3, col1}.
The function $G(\al, m)$ has been derived in \cite{corbos} for strongly correlated bosons when $Q(m)$ is the number of particles on a segment of ``length'' $m$.

\section{The state-vectors, the
Schur functions
and self-avoiding lattice walks}
\label{sec5}

\subsection{The Bethe state-vectors}

The present approach is based on the use of symmetric Schur functions \cite{macd} since this  is helpful for obtaining the correlation functions in the determinantal form \cite{b2, b3, bmnph}.

Let a set of strictly decreasing integers $\mu_k$, $1\leq k\leq N$, to constitute a \textit{strict partition}, i.e., $N$-tuple ${\bmu}=(\mu_1, \mu_2,\,\ldots\,, \mu_N)$ where
$M\geq \mu_1> \mu_2 > \, \dots\,> \mu_N \geq 1$. Since the operators $\si_n^{\pm}$ act on the states $\mid \uparrow \rangle_n$
and $\mid \downarrow \rangle_n$
according to (\ref{cor:lin411}), we define the state $|\bmu \rangle$ corresponding to $N$ flipped spins on the sites labelled by the ``coordinates'' $\mu_k$,
and the corresponding conjugate state $\langle \bnu |$:
\begin{equation}\label{conwf01}
|\bmu \rangle \equiv \begin{pmatrix}
\prod\limits_{k=1}^N \si_{\mu_k}^{-}\end{pmatrix} \mid
\Uparrow \rangle\,,\qquad\quad
\langle \bnu |\, \equiv \langle \Uparrow \mid \begin{pmatrix}
\prod\limits_{k=1}^N \si_{\nu_k}^{+}\end{pmatrix}\,,
\end{equation}
where $\mid\Uparrow\rangle
\equiv \bigotimes_{n=1}^{M} \mid \uparrow \rangle_n$. The states
(\ref{conwf01}) provide a complete orthogonal base:
\begin{equation}
\langle \bnu | \bmu \rangle =
\bdl_{\bnu \bmu} \equiv \prod\limits_{n=1}^N
\dl_{\nu_n \mu_n}\,.
\label{tuc3}
\end{equation}

The $N$-particles state-vectors $| \Psi({\bf u}_N)\rangle$ are chosen in the form of linear combinations of the states $| \bmu \rangle$ (\ref{conwf01}), \cite{bmnph, bmumn}:
\begin{equation}
\mid\!\Psi({\textbf u}_N)\rangle = \sum \limits_{\blad \subseteq \{{\CK}^N\}}
S_\blad ({\textbf u}^{2}_N)\,
|\bmu \rangle\,. \label{conwf1}
\end{equation}
The bold notations are adopted
in (\ref{conwf1}) (and hereinafter) for $N$-tuples of numbers like ${\bf u}^2\equiv (u^2_1, u^2_2, \dots , u^2_N)$ (or ${\bf u}_N^2$, to point out the number of elements). Summation in (\ref{conwf1}) goes over partitions ${\blad}=(\la_1, \la_2, \dots, \la_N)$ consisting of weakly decreasing non-negative integers: $\CK \geq \la_1\geq \la_2\geq \dots\geq
\la_N\geq 0$, where ${\CK}\equiv M-N$ is the number of
spins ``up''. The relationship between the \textit{parts} (i.e., elements) of $\blad$ and $\bmu$ is expressed as
\begin{equation}
\la_j=\mu_j+j-N-1\,, \qquad 1\le j\le N\,,
\label{eqnpart}
\end{equation}
or $\blad =\bmu - {\bdl}_N$,
where ${\bdl}_N$ is the ``staircase'' partition
\begin{equation}
{\bdl}_N \equiv (N, N-1, \dots, 2, 1)\,.
\label{strict}
\end{equation}
The coefficients in (\ref{conwf1}) are given by the {\it Schur functions}
$S_\blad$ defined by
the Jacobi--Trudi relation, \cite{bmumn}:
\begin{equation}
S_{\blad} ({\textbf x}_N)\,\equiv\,
\displaystyle{ S_{\blad} (x_1, x_2, \dots , x_N)\,\equiv\, \frac{\det(x_j^{\la_k+N-k})_{1\leq
j, k \leq N}}{\CV({\textbf x}_N)}}\,,
\label{sch}
\end{equation}
where $\CV ({\textbf x}_N)$ is the Vandermonde determinant
\begin{equation} \CV ({\textbf x}_N) \equiv
\det(x_j^{N-k})_{1\leq j, k\leq N}\,=\,
\prod_{1 \leq m< l \leq N}(x_l-x_m)\,.
\label{spxx1}
\end{equation}

With regard at (\ref {conwf01}),
the conjugate state-vectors are given by
\begin{equation}\label{conj}
\langle \Psi({\bf v}_N) |\,=\,\sum\limits_{\blad \subseteq \{{\CK}^N\}} \langle \bmu |\,
S_\blad ({\textbf v}^{-2}_N)\,.
\end{equation}
The scalar product of the states (\ref {conwf1}) and (\ref {conj}) takes the form:
\begin{equation}
\langle \Psi({\textbf
v}_N)\mid\!\Psi({\textbf u}_N)\rangle\,=\,
\sum_{\blad \subseteq \{{\CK}^N\}}S_\blad
({\textbf v}_N^{-2})S_\blad ({\textbf u}_N^2)\,,
\label{spxx3}
\end{equation}
where the ortho\-go\-na\-li\-ty  (\ref{tuc3}) is used.
Right-hand side of (\ref{spxx3}) is calculated by means of the Cauchy--Binet formula expressed through the Schur functions, \cite{gant}:
\begin{equation}
\sum_{\blad\subseteq \{L^N\}}S_{\blad}(\mathbf{x}_N) S_{\blad}(\mathbf{y}_N)= \frac{\det T_{L+N}(\textbf{x}_N, \textbf{y}_N)}{
{\CV} ({\textbf x}_N) {\CV} ({\textbf y}_N)}\,, \label{cauchy}
\end{equation}
where summation is over all partitions $\blad$
satisfying: $L\ge \la_1 \ge \la_2\ge \dots \ge \la_N \ge 0$. The matrix $T_{L+N}(\textbf{x}_N, \textbf{y}_N)$ $\equiv$ $(T_{i j}(\textbf{x}_N, \textbf{y}_N))_{1 \le i, j\le N}$ in (\ref{cauchy}) is given by the entries
\begin{equation}
T_{i j}(\textbf{x}_N, \textbf{y}_N)\equiv T_{i j}\equiv {\sf h}_{L+N} (x_i y_j)\,,\qquad {\sf h}_P(x) \equiv \frac{1-x^{P}}{1 - x}\,,
\label{cauchy1}
\end{equation}
where $P\in\BN$. Equations (\ref{cauchy}) and (\ref{cauchy1}) yield the scalar product (\ref{spxx3}):
\begin{equation}
\langle \Psi({\textbf
v}_N)\mid\!\Psi({\textbf u}_N)\rangle\,=\,\frac{1}{{\CV} ({\textbf v}^{-2}_N) {\CV} ({\textbf u}^{2}_N)}\,
\det \Bigl(\frac{1-(u_i/v_j)^{2M}}{1 - (u_i/v_j)^{2}}\Bigr)_{1\le i, j\le N}\,.
\label{spxx}
\end{equation}

Let us consider $N$-tuples ${\bf e}_{k}$, $1\le k\le N$, consisting of zeros except a unity at $k^{\rm th}$ place (say, from left). The Schur functions (\ref{sch}) labelled by a generic ${\blad}$ ($\la_1= {\CM}$ or $\la_N= 0$ for a non-generic ${\blad}$) respect the property:
\begin{equation}
\label{ratbe7272}
\sum_{k=1}^N S_{{\blad}\pm {\bf e}_{k}} ({\bf x}_N) = \Bigl(\sum_{k=1}^N x_k^{\pm 1} \Bigr) S_{{\blad}} ({\bf x}_N)\,.
\end{equation}

For a given ${\bf x}_N$, let us consider the set ${\cal S}\equiv \{S_{{\blad}} ({\bf x}_N)\}_{\blad \subseteq \{{\CK}^N\}}$ and subject its elements to the transformations ${\cal B}_N:\,{\blad} \rightarrow {\blad}\pm {\bf e}_{k}$, $k \in \{1, 2, \ldots, N\}$. The transformations ${\cal B}_N$ map arbitrary $S_{{\blad}} ({\bf x}_N)$ either to another element of the set or to zero. However, the transformations of the non-generic elements,
${\blad} \rightarrow {\blad}+ {\bf e}_{1}$
for ${\blad}=({\CM}, \ldots)$, or
${\blad} \rightarrow {\blad}- {\bf e}_{N}$ for ${\blad}=
(\ldots, 0)$, require a specification. Let us subject all $x_i\in {\bf x}_N$ to
\begin{equation}
\label{betheexp}
x_i^M=(-1)^{N-1}\,,\qquad 1\le i \le N\,.
\end{equation}
Then, the mapping of appropriate $S_{{\blad}} ({\bf x}_N)$ consists in transposition of the first/last column in the nominator of (\ref{sch}) ($j$ enumerates columns)
to the last/first position.
Thus, the set ${\cal S}_{{\cal B}} \equiv {\cal S}|_{(\ref{betheexp})}$ is mapped by ${\cal B}_N$ into itself, and we come to

\vskip0.3cm \noindent
\noindent{\bf Definition~1:\,}
\textit{Assume that $\si^{\pm}_n$ subjected to the periodicity
$\si^{\pm}_{n+M}=\si^{\pm}_n$, $\forall n\in {\cE}$, are used in $|\bmu\r$, $\l \bnu|$  \eqref{conwf01}. Then, \eqref{conwf1} and \eqref{conj} with the coefficients $S_{{\blad}} ({\bf x}_N) \in {\cal S}_{{\cal B}}$
are called $N$-particles Bethe state-vectors}.

\vskip0.3cm
Let us consider the exponential parametrization ${\textbf u}_N^{2}=e^{i
\bth_N}$, where $e^{{i\bth}_N}$ denotes $N$-tuple $(e^{i\ta_{1}}, e^{i\ta_{2}}, \ldots , e^{i\ta_{N}})$. It is directly verified that the Bethe state-vectors $\mid\!\Psi (e^{i{\bth }_N/2})\rangle$ introduced by {\sf Definition~1} are
the eigen-states of $H$ (\ref {cor:lin2}) and $S^z$ (\ref {cor:lin3}) on the periodic chain:
\begin{align}\label{egv}
\bigr(H_{\rm xx}- h S^z \bigl)\, \mid\!\Psi (e^{i{\bth }_N/2})\rangle\,=\, E_N({\bth }_N)
\mid\!\Psi (e^{i{\bth }_N/2})\rangle\,,\\
\label{egv1}
S^z \mid\!\Psi (e^{i{\bth }_N/2})\rangle\,=\,\Bigr( \frac{M}{2}-N\Bigl)
\mid\!\Psi (e^{i{\bth }_N/2})\rangle\,,
\end{align}
where $N$-tuple
$\bth_N \equiv (\ta_{1}, \ta_{2}, \dots , \ta_{N})$ is defined, and \eqref{betheexp} is nothing but the set of the famous Bethe equations in the exponential form \cite{col1} for the $XX$ chain: $e^{i M\ta_j}=(-1)^{N-1}$, $1\le j \le N$. The eigen-energy $E_N(\bth)$ (\ref{egv}) is equal to
\begin{equation}
E_N(\bth)\,=\,-\frac{hM}2 + \sum_{j=1}^N
\ep (\ta_j)\,,\qquad
\ep (\ta_j) \equiv h-
\cos\ta_j\,,
\label{egen}
\end{equation}
where
\begin{equation}
\theta_j = \frac{2\pi
}{M}
\begin{pmatrix}
\displaystyle{
I_j-\frac{N+1}{2}} \end{pmatrix}\,,
\quad 1\le j \le N\,, \label{besol}
\end{equation}
and $I_j$ are integers, $M \geq I_1>I_2> \dots>I_N\geq 1$, constituting $N$-tuple ${{\bf{I}}}_N = (I_1, I_2, \dots, I_N)$.
The \textit{ground state} solution is given by (\ref{besol}) with ${{\bf{I}}}_N$
substituted by $\bdl_N$ (\ref{strict}):
\begin{equation}\label{grstxx}
\theta^{\,\rm g}_j \equiv \frac{2\pi
}{M}\begin{pmatrix} \displaystyle{\frac{N+1}{2} -j} \end{pmatrix}\,, \quad
1\le j \le N\,.
\end{equation}
Useful relations result from
(\ref{conwf1}), (\ref{conj}),  (\ref{egv}):
\begin{equation}
\begin{aligned}
\langle {\bmu}_N | e^{-\be H} |\Psi(e^{i{\bth}_N/2})
\rangle & =\, e^{-\be E_N ({\bth}_N) }\,
S_{\blad_N} (e^{i{\bth}_N}), \\[0.3cm]
\langle \Psi(e^{-i{\bth}_N/2}) | e^{-\be H} |
{\bmu}_N \rangle &=\, e^{-\be E_N ({\bth}_N) }\,
S_{\blad_N} (e^{-i{\bth}_N})\,.
\label{ratbe91}
\end{aligned}
\end{equation}

Let us introduce $\CN^2({\textbf u}_N) \equiv \langle \Psi({\textbf
u}_N)\mid\!\Psi({\textbf u}_N)\rangle$ for the scalar product (\ref{spxx}) of the states (\ref{conwf1}) at ${\textbf v}_N={\textbf u}_N$. Then the square of the norm
$\CN^2(e^{i{\bth}_N/2})$ parameterized by solution to the Bethe equations (\ref{betheexp})
takes the form due to (\ref{cauchy}) and (\ref{spxx}), \cite{bmumn}:
\begin{equation}
\CN^2 (e^{i{\bth}_N/2})\,=\,
\displaystyle{
\frac{M^N}{| \CV (e^{i{\bth}_N}) |^2}
=\frac{M^N}{\prod \limits_{1\leq m<l\leq
N} 2 (1-\cos \frac{2
\pi}{M}(I_l-I_m))}}\,. \label{normxx}
\end{equation}
Decomposition of unity is of the form:
\begin{equation}
{\BI}\,=\,\sum\limits_{\{{\bth}_N\}}
\CN^{-2} (e^{i{\bth}_N/2})
\mid\!\Psi (e^{i{\bth}_N/2})\rangle \langle
\Psi (e^{i{\bth}_N/2})\!\mid\,, \label{field7}
\end{equation}
where (\ref {conwf1}) and (\ref {conj}) are taken into account, $\CN^{2} (e^{i{\bth}_N/2})$ is given by (\ref{normxx}), and
summation is over all independent solutions to
(\ref{betheexp}).

\subsection{The Schur functions, self-avoiding lattice paths, and boxed plane partitions}

\begin{figure}[h]
\centering
\includegraphics
{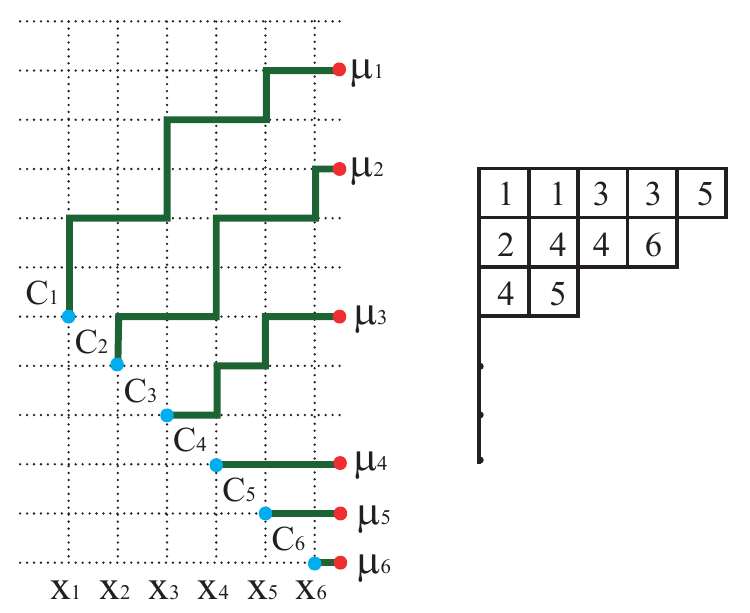}
\caption{A nest $\cal{C}$ of $N=6$ lattice paths and semi-standard tableau ${\sf {T}}$ of shape $\blad=(5, 4, 2, 0, 0, 0)$.}
\label{fig:f5}
\end{figure}

The Schur functions $S_{\blad}(\mathbf{x}_N)$ (\ref{sch}) are in one-to-one correspondence with the semi-standard Young tableaux \cite{macd}, and
they admit an interpretation in terms of self-avoiding lattice walks. A semi-standard Young tableau ${\sf {T}}$ of shape ${\blad}$
is a diagram possessing $\lambda_i$ cells in $i^{\rm th}$ row ($i=1, \ldots, N$). The cells are filled with positive integers $n\in \BN^+$ weakly increasing along rows and strictly increasing along columns (right-hand side of Fig.~\ref{fig:f5}). A   \textit{nest of self-avoiding lattice paths} $\cal{C}$ (left-hand side of Fig.~\ref{fig:f5}) consists of paths counted from the top of ${\sf {T}}$ and going from points $C_i=(i, N+1-i)$ to points ($N, \mu_i=\la_i+N+1-i$). An $i^{\rm th}$ lattice path
makes $\la_i$ upward steps, and it encodes $i^{\rm th}$ row of the tableau. The number $l_j$ of upward steps along the line $x_j$ coincides with the number of occurrences of $j$ in ${\sf {T}}$. Then, $S_{\blad} (\mathbf{x}_N)$ (\ref{sch}) corresponding to ${\sf {T}}$ of shape ${\blad}$ takes the form:
\begin{equation}\label{schrepr}
S_{\blad} (\mathbf{x}_N) = \sum_{\{\cal{C}\}} \prod_{j=1}^{N}
x_{j}^{l_j},
\end{equation}
where summation is over all admissible nests $\cal{C}$. Let us notice how (\ref{schrepr}) enables to obtain (\ref{ratbe7272}). The set of all
semi-standard Young tableau of shapes
${{\blad} \pm {\bf e}_{k}}$, $1\le k\le N$, is characterized by the volume $|{\blad}|\pm 1$. Since $\sum_{i=1}^{N} l_i = |{\blad}|$, one concludes that (\ref{ratbe7272}) is valid.
The representation (\ref{schrepr}) naturally arises in quantum models soluble by the Quantum Inverse Scattering Method \cite{KBI2}.
The value $S_{\blad} ({\bf 1}_N) \equiv S_{\blad} (1, 1, \ldots, 1)$ gives the number of nests of self-avoiding lattice paths, and it is equal to
\begin{equation}\label{numbpaths1}
S_{\blad} ({\bf 1}_N) = \prod_{1\leq j<k\leq N}\frac{\lambda_j - j - \lambda_k+k}{k-j} = \prod_{1\leq j<k\leq N} \frac{\mu_j - \mu_k}{k-j}\,.
\end{equation}

\begin{figure}[h]
\centering
\includegraphics
{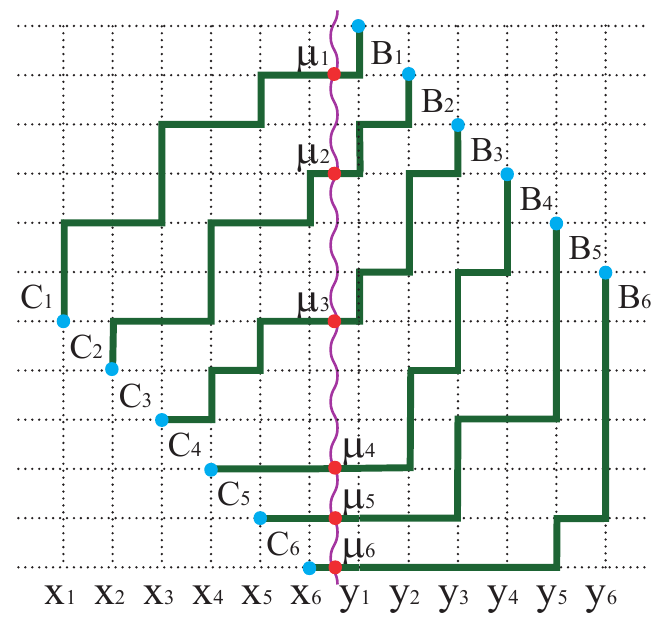}
\caption{\textit{Watermelon} as the nest of lattice paths at  ${\cal M}=6$, $N=6$.}
\label{fig:f6}
\end{figure}

Let us consider the nest of $N$ self-avoiding lattice paths with equidistantly arranged start and end points $C_l$ and $B_{l}$, respectively ($1\le l \le N$). Only upward and rightward steps are allowed for the path in the nest so that an $l^{\rm th}$ one is contained within the rectangle whose lower left and upper right vertices are $C_l$ and $B_l$, respectively. Besides, the total number ${\CM}=M-N$ of upward steps and the total number $N$ of rightward steps are the same for each path belonging to the nest. Then the nest described is called \textit{watermelon}
(see Fig.~\ref{fig:f6}).

Watermelon can be viewed as a `fusion' (`sewing') of the nest of paths ${\cal C}$ and of a \textit{conjugate} nest of paths ${\cal B}$
along the points on the `dissection' line (wavy line in Fig.~\ref{fig:f6}). The partition $\bmu$ determines the ordinates of the points $(N, \mu_l)$, $M \ge \mu_1 > \mu_2 > \cdots > \mu_N \ge 1$, which are the end points of the nest ${\cal C}$ and which must coincide with those characterizing the conjugate nest ${\cal B}$. For instance, a typical watermelon in Fig.~\ref{fig:f6} is given by the nest ${\cal C}$ (see Fig.~\ref{fig:f5}) fused with a conjugate nest ${\cal B}$ which can be restored from Fig.~\ref{fig:f6}, \cite{bmumn}. The Schur function corresponding to the conjugate nest of $N$ self-avoiding paths is
\begin{equation}\label{rsf}
S_{\blad} ({\bf y}_N) \equiv
S_{\blad} (y_1, y_2, \ldots, y_N) = \sum_{\{\cal{B}\}} \prod_{r=1}^{N}
y_{r}^{{\cal M}-b_r},
\end{equation}
where $b_r$ is the number of upward steps along $y_{r}$, and summation is over all nests $\cal{B}$.

Under the $q$-parametrization
\begin{equation} \label{rep21}
\textbf{v}^{-2}={\bf q}_N\equiv (q, q^2, \dots, q^N)\,,\qquad
\textbf{u}^2 = {{\bf q}_N}/q
\,,
\end{equation}
the scalar product (\ref{spxx3})
takes the form:
\begin{equation}
\langle \Psi(\textbf{q}^{-1/2}_N)|
\Psi((\textbf{q}_N/{q})^{1/2})
\rangle\,=\,
\sum_{\blad \subseteq \{{\CK}^N\}} S_{\blad} ({\bf q}_N) S_{\blad} \Bigl(\frac{{\bf q}_N
}{q}\Bigr)\,.
\label{spxx03}
\end{equation}
Then, the number of watermelons characterized by the points $C_l$ and $B_{l}$ ($1\le l \le N$)
is given by (\ref{spxx03}) at $q\to 1$:
\begin{align}
\lim_{q\to 1}\,
\langle \Psi(\textbf{q}^{-1/2}_N)|
\Psi((\textbf{q}_N/{q})^{1/2})
\rangle
 &=\, \displaystyle{
\sum_{\blad \subseteq \{{\CK}^N\}} S_{\blad} (\mathbf{1}_N) S_{\blad} (\mathbf{1}_N) }
\nonumber \\
\label{schrepr5}
& = \sum_{\bmu \subseteq \{{M}^N\}}
\Bigl(\sum_{\{\cal{C}\}_{\bmu}} 1\Bigr) \Bigl(\sum_{\{\cal{B}\}_{\bmu}} 1\Bigr)\,,
\end{align}
where the notations  $\{\cal{C}\}_{\bmu}$ and $\{\cal{B}\}_{\bmu}$ are to stress that the summations are over the nests characterized by specific $\bmu$ (more on the graphical interpretation in \cite{bmumn}).

\begin{figure}[h]
\center
\includegraphics [scale=0.5]{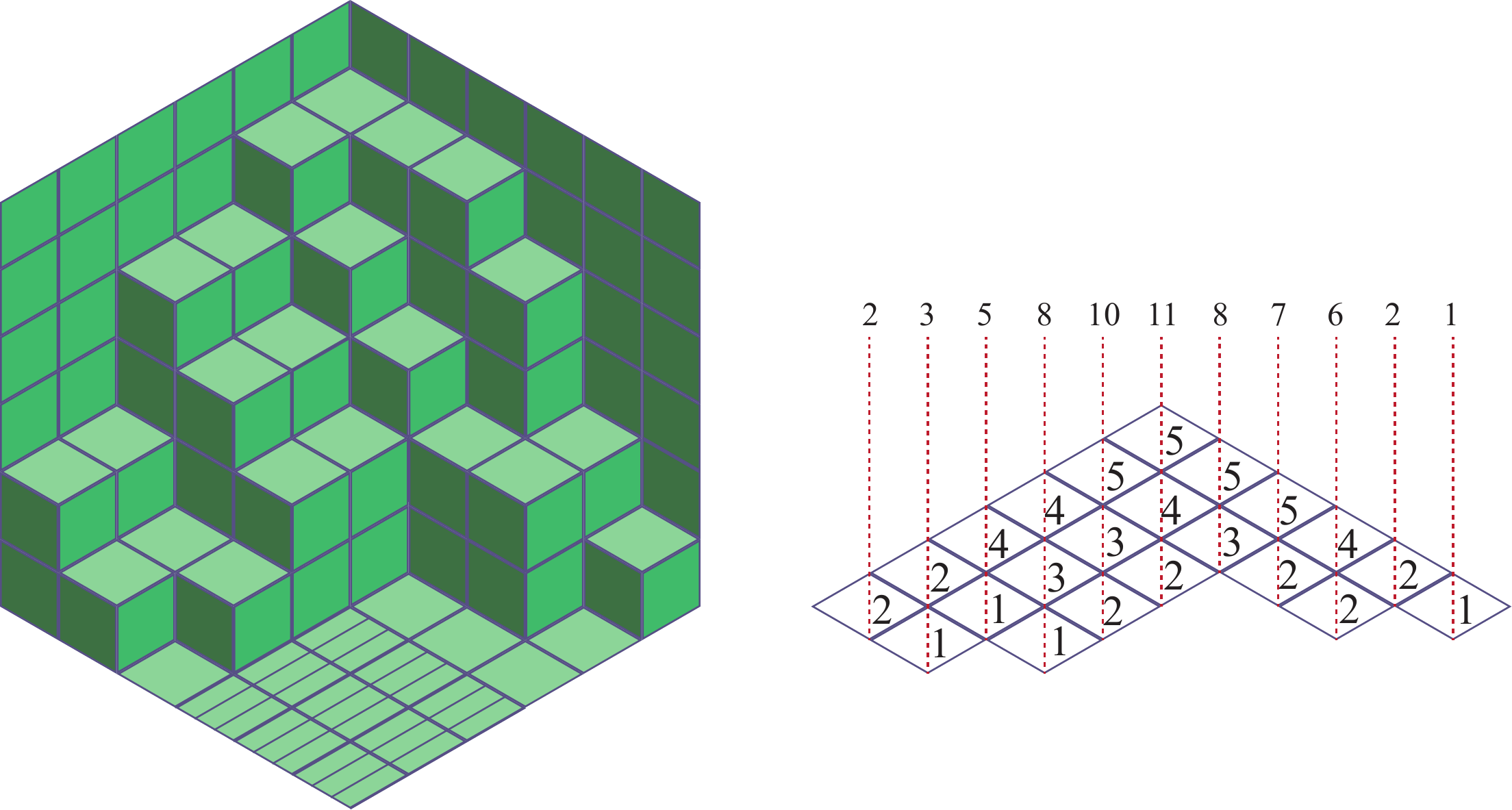}
\caption{Plane partition equivalent to watermelon in Figure \ref{fig:f6} ($|\blad|=11$).}
\label{fig:ff9}
\end{figure}

A \textit{boxed plane partition} ${\bpi}$ is an array $({\pi}_{ij})_{i, j\ge 1}$ of non-negative
integers that satisfy ${\pi_{ij} \geq \pi_{i+1, j}}$ and ${\pi_{ij} \geq \pi_{i, j+1}}$ for all ${i, j \geq 1}$, \cite{bres, stan1}. A boxed plane partition is
contained in ${L\times N\times M}$ box, if ${\pi_{ij} \leq M}$
for all ${i}$ and ${j}$, and ${\pi_{ij}=0}$, whenever ${i>L}$ or ${j>N}$. Plane
partitions are interpreted as stacks of unit cubes so that the height of a stack at point $(i, j)$ is ${\pi_{i j}}$ (left-hand side of Fig.~\ref{fig:ff9}).
The trace of $s^{\rm th}$ diagonal of plane partition
counted from left down corner
is $\tr_s {\bpi}\equiv \sum_{N+j-i=s} \pi_{i j}$, $1\le s\le 2N-1$. The volume of ${\bpi}$ is $|{\bpi}| = \sum_{s=1}^{2N-1} {\tr}_s {\bpi}$. The plane partition and the corresponding array are depicted in Fig.~\ref{fig:ff9}. There exists bijection between the watermelon configuration of self-avoiding lattice paths (Fig.~\ref{fig:f6}) and the plane partitions (Fig.~\ref{fig:ff9}), \cite{bmumn}. The bijection is such $\tr_N {\bpi}=|{\blad}|$ (all traces are provided in Fig.~\ref{fig:ff9}).

The generating functions of boxed plane partitions arise from the correlation functions of the $XX$ model \cite{b2, b3, bmnph}, the quantum phase model \cite{statm}, the four-vertex model \cite{bmjpa}.

\section{Norm-trace generating function of plane partitions as form-factor of $\exp{\cal Q}$}
\label{sec53}
\subsection{Flipped spins and plane partitions with columns of fixed heights}

The commutation rule
\begin{equation}
\label{cor:lin551}
e^{\cQ}\,\si^\pm_{k}\,=\,e^{\mp\al_k} \si^\pm_{k}\,
e^{\cQ}
\end{equation}
and ${\cQ}\mid\Uparrow\rangle = 0$ allow us to obtain
the average of $e^{\cQ}$ over
`off-shell' (i.e., arbitrarily parameterized) $N$-particles
states \eqref{conwf1} and \eqref{conj}:
\begin{equation}
\langle \Psi({\bf v}_N)\mid e^{\cQ}\mid \!\Psi({\textbf u}_N)\rangle =
{\cal P}_{\CK}({\textbf v}^{-2}_N, {\textbf
u}^2_N, {\bf a}_M)\,,
\label{cor:liin552}
\end{equation}
where
\begin{equation}
{\cal P}_{\CK}({\textbf v}^{-2}_N, {\textbf
u}^2_N, {\bf a}_M)
\equiv
\sum_{\blad \subseteq \{{\CK}^N\}}S_\blad
({\textbf v}^{-2}_N)S_\blad ({\textbf
u}^2_N) \prod_{i=1}^N e^{\al_{\mu_i}}
\label{cor:lin552}
\end{equation}
is the sum depending on the elements of $M$-tuple
${\bf a}_M \equiv (\al_{1},
\al_{2}, \ldots ,
\al_{{M}})$, while the parts of $\blad$ and $\bmu$ are related according to (\ref{eqnpart}).
The use of the generic Cauchy--Binet formula
\cite{gant} leads to

\vskip0.3cm \noindent
\noindent{\bf Proposition~1:\,}
\textit{The sum ${\cal P}_{\CK} ({\bf v}^{-2}_N, {\bf
u}^2_N, {\bf a}_M)$ \eqref{cor:lin552} parameterized by $M$-tuple ${\bf a}_M$
and by the arguments ${\bf v}^{-2}_N$ and ${\bf u}^2_N$ of the Schur functions
admits the determinantal representation}:
\begin{equation}
{\cal P}_{\CK}({\textbf v}^{-2}_N, {\textbf
u}^2_N, {\bf a}_M)\,=\, \displaystyle{
\frac{1}{{\CV} ({\textbf
u}^2_N){\CV} ({\textbf v}^{-2}_N)} \det
\begin{pmatrix} \displaystyle{ \sum\limits_{n=1}^{M} e^{\al_n}\,
\Bigl(\frac{u_i^{2}}{v_j^{2}} \Bigr)^{n-1} } \end{pmatrix}_{1\le i, j \le N}}\,,
\label{cor:lin553}
\end{equation}
\textit{where the Vandermonde determinant \eqref{spxx1} is used}.
\vskip0.3cm

Off-shell $N$-particles average of the projector ${\varPi}_{\bf k}$ defined in (\ref{cor:lin611}) arises from the series representation (\ref{cor:liin552}) provided that
the notations for $l$-tuples ${\bf k}_l\equiv (k_1, k_2, \ldots, k_l)$ and the ``reversed'' one $\bar{{\bf k}}_l \equiv (k_l, k_{l-1}, \ldots, k_1)$ are adopted:
\begin{align}
\langle \Psi({\bf v}_N)\mid
{\varPi}_{\bf k}
\mid\! \Psi({\textbf u}_N)\rangle\,=\,
\lim\limits_{\{\al_k \to 0\}}\,\cd^{\,l}_{\al_{k_1}, \al_{k_2},\dots ,
\al_{k_{l}}} \nonumber\\
\times\,\langle \Psi({\bf v}_N)\mid e^{\cQ}\mid\!\Psi({\textbf u}_N)\rangle\,=\,{\w {\cal P}}_{\CK}({\textbf v}^{-2}_N, {\textbf
u}^2_N, {\bf k}_l)\,,
\label{5544}
\end{align}
where the tilded notation ${\w {\cal P}}_{\CK} ({\textbf v}^{-2}_N, {\textbf u}^2_N, {\bf k}_l)$ implies the sum
\begin{equation}
\label{55441}
{\w {\cal P}}_{\CK}({\textbf v}^{-2}_N, {\textbf
u}^2_N, {\bf k}_l)\,\equiv\,
\sum_{\w{\blad} \subseteq \{{\CK}^N\}}S_{\w{\blad}}
({\textbf v}^{-2}_N) S_{\w{\blad}} ({\textbf
u}^2_N)\,.
\end{equation}
Summation in (\ref{55441}) goes over $\w{\blad}\equiv \w{\blad}_N = \w{\bmu}_N - {\bdl}_N$, where $\w{\bmu}_N$ is a strict partition such that its $l$ non-consecutive parts coincide with the elements of $l$-tuple $\bar{{\bf k}}_l$, and ${\bdl}_N$ is given by (\ref{strict}).

The average of ${\varPi}_{\bf k}$
under the $q$-parametrization  (\ref{rep21}) arises from (\ref{5544}) and (\ref{55441}):
\begin{equation}
\langle \Psi(\textbf{q}^{-1/2}_N)\mid {\varPi}_{\bf k} \mid\!
\Psi((\textbf{q}_N/{q})^{1/2})
\rangle\,=\,{\w {\cal P}}_{\CK}
\Bigl(\textbf{q}_N, \frac{\textbf{q}_N}{q}, {\bf k}_l\Bigr)
\label{cor:lenn5571} \,.
\end{equation}
Equation (\ref{cor:lenn5571}) in the case ${\bf k}_l = {\bf l}
\equiv (1, 2, \ldots, l)$ takes the form:
\begin{equation}
\langle \Psi(\textbf{q}^{-1/2}_N)\mid \prod_{i=1}^{l} {\sf q}_{i} \mid\!
\Psi((\textbf{q}_N/{q})^{1/2})
\rangle\,=\,{\w {\cal P}}_{\CK}
\Bigl(\textbf{q}_N, \frac{\textbf{q}_N}{q}, {\bf l}\Bigr)
\label{cor:lin5571} \,,
\end{equation}
so that $\w{\bmu}_N$ and $\w{\blad}_N = \w{\bmu}_N - \bdl_N$ in (\ref{55441}) are concretized as follows:
\begin{align}
\w{\bmu}_N & = (\mu_1, \mu_2, \ldots, \mu_{N-l}, l, l-1, \ldots, 1)\,,
\label{cor:lin558} \\
\w{\blad}_N & = (\la_1, \la_2, \ldots, \la_{N-l}, 0, 0, \ldots, 0)\,,
\label{cor:lin5581}
\end{align}
and summation in ${\w {\cal P}}_{\CK}$ (\ref{55441}) is over $\CK\geq \la_1\geq \la_2\geq \dots\geq
\la_{N-l}\geq 0$.

According to (\ref{schrepr5}), right-hand side of (\ref{cor:lin5571}) provides the generating function of the number of watermelons depicted in Fig.~\ref{fig:f6}:
\begin{align}
\langle \Psi(\textbf{1}_N)\mid \prod_{i=1}^{l} {\sf q}_{i}\mid\! \Psi(\textbf{1}_N)
\rangle &=\,{\w {\cal P}}_{\CK}
(\textbf{1}_N, \textbf{1}_N, {\bf l}) \nonumber \\
& = \lim_{q\to 1}
\sum_{\tilde\blad \subseteq \{{\CK}^N\}} S_{\w{\blad}} (\textbf{q}_N)
S_{\w{\blad}}\Bigl(
\frac{ \textbf{q}_N} {q}\Bigr)
= \sum_{\tilde\blad \subseteq \{{\CK}^N\}} S_{\w\blad} (\textbf{1}_N)
S_{\w\blad} (\textbf{1}_N)
\label{cor:lin5572} \,.
\end{align}
Indeed, $S_{\w\blad} ({\bf 1}_N)$ corresponds to the paths  connecting equidistant points $C_i=(i, N+1-i)$ with non-equidistant ones ($N, \w\mu_i$), where $\w\mu_i$ are given by (\ref{cor:lin558}). The nest in Fig.~\ref{fig:f5} is just depicted for $\w\bmu$ (\ref{cor:lin558}) since an $i^{\rm th}$ path makes $\la_i \in \blad_{N-l}\equiv (\lambda_1, \lambda_2, \ldots, \lambda_{N-l})$ steps upwards
at $1\le i\le N-l$, while only rightward steps are allowed at $N-l+1\le i\le N$ ($l=3$ in Fig.~\ref{fig:f5}).
The following identity is respected by $S_{\w\blad} ({\bf 1}_N)$ due to (\ref{numbpaths1}) and (\ref{cor:lin5581}):
\begin{equation}
S_{\w\blad} ({\bf 1}_N) =
S_{\blad_{N-l}} ({\bf 1}_N)\times \prod_{k=1}^{N-l}\prod_{j=N-l+1}^{N}
\frac{\lambda_k+j-k} {j-k}\,.
\label{numbpaths}
\end{equation}

In the case of $l=0$, Eq.~(\ref{cor:lin5572}) is reduced to (\ref{schrepr5}).
Therefore, $\langle \Psi(\textbf{1}_N) | \Psi(\textbf{1}_N)
\rangle$ is given by the sum ${\cal P}_{\CK}({\bf 1}_N, {\bf 1}_N, {\bf 0}_{M})$ (\ref{cor:lin552}) equal to the number of such watermelons that upward steps are allowed for
all paths (including, in comparison with Eq.~(\ref{cor:lin5572}), the paths from $(N-l+1)^{\rm th}$ to $N^{\rm th}$). The number ${\cal P}_{\CK}({\bf 1}_N, {\bf 1}_N, {\bf 0}_{M})$ is also interpreted as the number $A (N, N, M-N)$ of plane partitions in $N\times N \times (M-N)$ box (see Figure \ref{fig:ff9}), \cite{bmumn}:
\begin{align}
\nonumber
\langle \Psi(\textbf{1}_N)\mid\! \Psi(\textbf{1}_N)
\rangle &=\,
{\cal P}_{\CK}({\bf 1}_N, {\bf 1}_N, {\bf 0}_{M}) \\ \label{ratbe794}
&= A (N, N, M-N) = \prod_{k=1}^{N} \prod_{j=1}^{N} \frac{M-N+k+j-1}{k+j-1}\,.
\end{align}

As far as the mapping between the watermelon configurations and the plane partitions is concerned, the watermelons characterized by $\w{\bmu}_N$ (\ref{cor:lin558}) and $\w{\blad}_N$ (\ref{cor:lin5581}) are mapped to such stacks of cubes that $l\times l$ square on the bottom of $N\times N\times {\CM}$ box remains empty. The specific watermelon in Figure \ref{fig:f6} is characterized by
$\mu_4=3$, $\mu_5=2$, $\mu_6=1$, and $\la_4=\la_5=\la_6=0$.
The dashed $3\times 3$ square is shown in Figure \ref{fig:ff9}. It is forbidden for the cubes constituting the specific stacks to occupy the dashed square.
Therefore, ${\w {\cal P}}_{\CK}
(\textbf{1}_N, \textbf{1}_N, {\bf l})$ (\ref{cor:lin5572}) enumerates the plane partitions restricted additionally by an ``excluded'' part of the bottom surface. Generally, ${\w {\cal P}}_{\CK}
(\textbf{1}_N, {\textbf{1}_N}, {\bf k}_l)$ corresponding to (\ref{cor:lin5571}) enumerates the plane partitions with $l$ columns of prescribed height in one-to-one correspondence with parts of $\bar{{\bf k}}_l$.

\subsection{Norm-trace generating function}

The \textit{norm-trace generating function} ${G}(N, N, {\CM} |\,q, \ga)$, i.e., the generating function of plane partitions with unbounded parts and with fixed height
of the main diagonal in a box of height $\CM$ and with bottom of size $N\times N$ has been derived in \cite{st} and generalized in \cite{gan}. The determinantal representation for  ${G}(N, N, {\CM} |\,q, \ga)$ has been derived for the model of strongly correlated bosons
\cite{statm}. The determinantal formula for the generating function of plane partitions with fixed heights of several diagonals has been obtained by means of the four-vertex model in inhomogeneous field \cite{bmjpa}.

The norm-trace generating function ${G}(N, N, {\CM} |\,q, \ga)$ for the Heisenberg $XX$ chain arises from Eqs.~(\ref{cor:liin552}) and (\ref{cor:lin552})
under the $q$-parametrization (\ref{rep21}). Indeed, let us
consider the linear parametrization of ${\bf a}_M$ and specify $\al_n$ so that $e^{\al_n} = \ga^n$, $0<{\ga}\le 1$. We shall use $\langle e^{\cQ (\ga)} \rangle_{N, q}$ to denote the corresponding $q$-parameterized average (\ref{cor:liin552}):
\begin{equation}
\langle e^{\cQ (\ga)}
\rangle_{N,q}\,\equiv\,\langle \Psi(\textbf{q}^{-1/2}_N)\mid e^{\cQ (\ga)}\mid\!
\Psi((\textbf{q}_N/{q})^{1/2})
\rangle\,.
\label{ratbee707}
\end{equation}

One formulates the following

\vskip0.3cm \noindent
\noindent{\bf Proposition~2:\,}
\textit{The determinantal representation for
the norm-trace generating function of plane partitions with fixed height of their diagonal parts in a box of height $\CM$ and bottom of size $N\times N$ is given}:
\begin{align}
{G}(N, N, {\CM} |\,q, \ga) & =
\ga^{\frac{-N}{2}(N+1)}\,\langle e^{\cQ (\ga)}
\rangle_{N,q}
\nonumber \\
&=\,\displaystyle{
\frac{\det\big(
{\sf h}_M( {\ga}\, q^{i+j-1})\big)_{1\le i, j \le N}}{{\CV} (\textbf{q}_N/{q}) {\CV} (\ga\,\textbf{q}_N)}}\,,
\label{ratbe707}
\end{align}
\textit{where ${\sf h}_M$ is defined by} (\ref{cauchy1}).

\vskip0.3cm \noindent
\noindent{\bf Proof:\,}
First of all, one obtains from (\ref{cor:lin552}) and (\ref{cor:lin553}):
\begin{align}
\ga^{\frac{-N}{2}(N+1)}\,
\langle e^{\cQ (\ga)}
\rangle_{N,q} &=\,
\sum_{\blad \subseteq \{{\CK}^N\}} \gamma^{|{\blad}|} S_{\blad} \Bigl(\frac{\textbf{q}}{q}\Bigr)
S_{\blad} (\textbf{q})
\label{cor:lin559} \\
&=\,\displaystyle{
\frac{1}{{\CV} (\textbf{q}_N/{q}) {\CV} (\ga\,\textbf{q}_N)} \det
\begin{pmatrix} \displaystyle{ \sum\limits_{n=0}^{M-1} \ga^{n}
q^{n(i+j-1)} } \end{pmatrix}_{1\le i, j \le N}}\,,
\label{cor:lin5591}
\end{align}
where $|\blad| = \sum_{i=1}^{N}{\la_i}$ is the \textit{weight} of $\blad$. The relation $|\bmu|=|\blad| + \frac{N}{2} (N+1)$ is used for obtaining
(\ref{cor:lin559}). The homogeneity property $\ga^{|\blad|} S_{\blad} (\textbf{q}) = S_{\blad} (\ga\,\textbf{q})$ is used to pass from (\ref{cor:lin559}) to (\ref{cor:lin5591}).
The series in right-hand side of (\ref{cor:lin559}) is by definition the norm-trace generating function of plane partitions with fixed height of their diagonal parts in $N\times N\times \CM$ box, and therefore
the statement (\ref{ratbe707}) for ${G}(N, N, {\CM} |\,q, \ga)$ is valid due to the determinanal formula (\ref{cor:lin5591}), \cite{statm}. Equation (\ref{ratbe707}) at ${\ga}=1$ gives the determinantal formula for the generating function of boxed plane partitions in $N\times N\times {\CM}$ box:
\begin{equation}
\label{cor:llin5591}
\lim_{q\to 1}\,{G}(N, N, {\CM} |\,q, 1) = A (N, N, {\CM}) \,,
\end{equation}
where the number of plane partitions $A (N, N, {\CM})$ is given by (\ref{ratbe794}) (MacMahon formula, \cite{bres}). $\Box$

Assume that the approximation ${\sf h}_M(x) \simeq (1-x)^{-1}$ is valid at $|x|\le 1$ and large enough $M$. Then, one obtains from (\ref{ratbe707}):
\begin{align}
\lim_{M\to\infty}
{G}(N, N, {\CM} |\,q, \ga) & =\,
\frac{\det \begin{pmatrix} \bigl(1- \ga
q^{i+j-1}\bigr)^{-1} \end{pmatrix}_{1\le i, j \le N}}
{{\CV} (\textbf{q}_N/{q}) {\CV} (\ga\,\textbf{q}_N)}
\label{cor:lin5596}
\\
&=\,\displaystyle{
\prod_{i=1}^{N} \prod_{j=1}^{N} \frac{1}{1- \ga
q^{i+j-1} }}\,.
\label{cor:lin5597}
\end{align}
Evaluation of the
Cauchy-type determinant in right-hand side of (\ref{cor:lin5596}) leads to the double product  (\ref{cor:lin5597}), which is nothing but the norm-trace generating function of plane partitions with unbounded height \cite{statm}. Further, one obtains from (\ref{cor:lin5597}) the limiting expression
\begin{equation}\nonumber
\lim_{N/M\ll 1, N \to\infty}
{G}(N, N, {\CM} |\,q, \ga) \,=\,\displaystyle{
\prod_{n=1}^{\infty}  \frac{1}{(1- \ga
q^{n})^n }}\,,
\end{equation}
which is related with the partition function of the five-dimensional supersymmetric Yang-Mills theory \cite{jap}.

\section{The transition amplitude and random turns walks of vicious walkers}
\label{50}
\subsection{Multi-particles transition amplitude}
\label{s341}

One-dimensional random walks of \textit{vicious walkers} who annihilate one another whenever they meet at the same lattice site attract attention after \cite{fish}, and so-called {\it lock step}, \cite{forr1}, and {\it random turns} models, \cite{forr2, forr},
are distinguished. Suppose that there are $N$ walkers on a one-dimensional lattice. In the random turns model only a single randomly chosen walker moves
at each tick of a clock
to one of closest sites while the rest are staying.
It has been proposed in \cite{b1, b11} to interpret random movements in the random turns model as transitions between spin ``up'' and ``down'' states of the Heisenberg $XX$ chain.

\begin{figure}[h]
\centering
\includegraphics[scale=1.0]{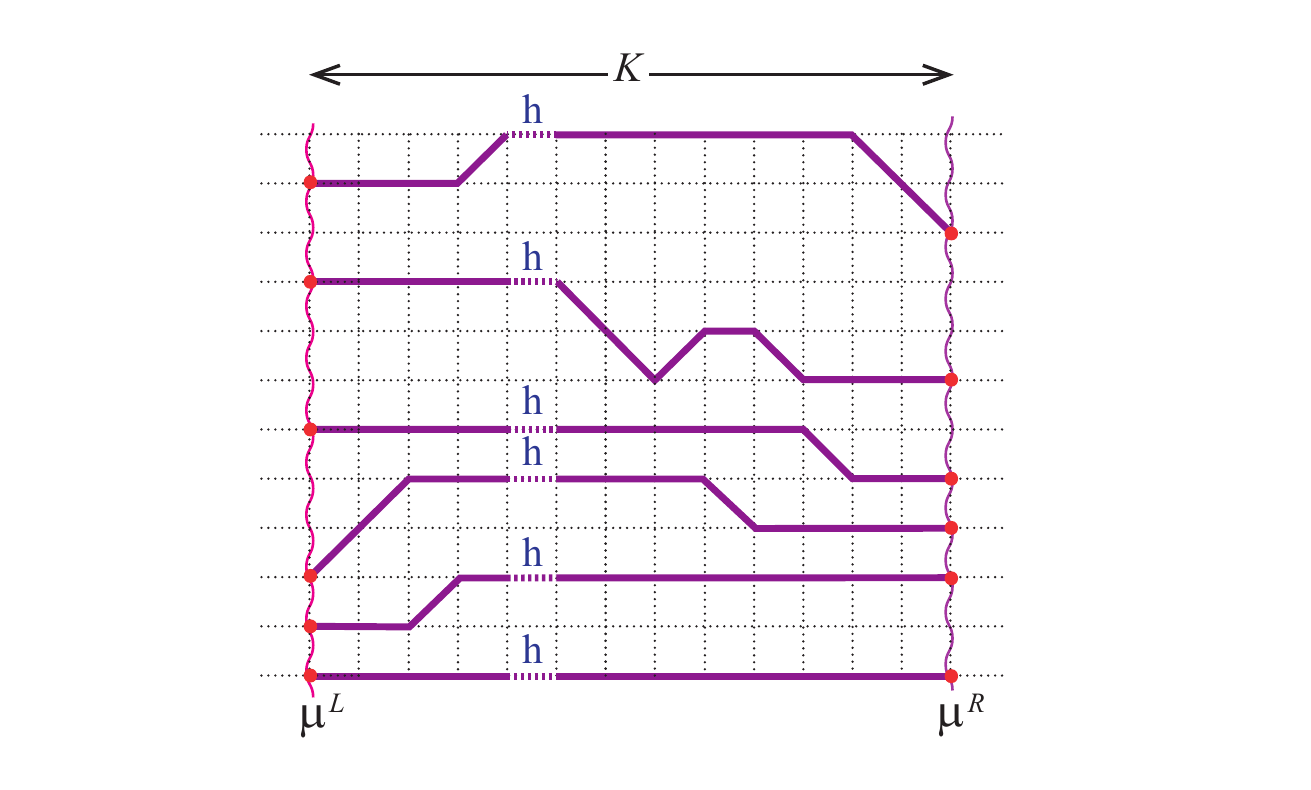}
\caption{Random turns vicious walkers.}
\label{fig:f1}
\end{figure}

The generating function of the lattice trajectories of $N$ random turns vicious
walkers (a typical example in Fig.~\ref{fig:f1})
is given by $N$-particles \textit{transition amplitude}
corresponding to the $XX$ Heisenberg model described by the Hamiltonian (\ref{cor:lin2}):
\begin{equation}
G_{{\bmu^L}; {\bmu^R}}(\be) \equiv \langle {\bmu^L} |\, e^{- \be H_{\rm xx}+ \be h S^z} |\, {\bmu^R} \rangle\,,
\label{mpcf}
\end{equation}
which is parameterized by parts of ${\bmu^L} \equiv (\mu^L_1, \mu^L_2, \dots, \mu^L_N)$ and ${\bmu^R}\equiv(\mu^R_1, \mu^R_2, \dots, \mu^R_N)$ interpreted as
initial and final positions of the walkers. The representation
(\ref{mpcf}) is re-expressed as follows:
\begin{align}
G_{{\bmu^L}; {\bmu^R}}(\be) &=\,
e^{\be h (\frac{M}2 - N)}
G^{\,0}_{{\bmu^L}; {\bmu^R}} (\be) \,,
\label{mpcf21} \\[0.2cm]
G^{\,0}_{{\bmu^L}; {\bmu^R}} (\be) & \equiv
\langle {\bmu^L} |\, e^{- \be H_{\rm xx}} |\, {\bmu^R} \rangle\,,
\label{mpcf2}
\end{align}
provided that the commutation relation
\begin{equation}
\nonumber
e^{\be h S^z} \si^\pm_n\,=\,
e^{\pm\be h} \si^\pm_n e^{\be h S^z}
\end{equation}
is accounted for together with $S^z \mid \Uparrow \rangle = \frac M2 \mid \Uparrow \rangle$. The exponential factor in right-hand side of (\ref{mpcf21}) is due to coupling of the spin chain to the homogeneous magnetic field, and the corresponding exponent is proportional to the eigen-value (\ref{egv1}) of the total spin.

The present approach to $G_{{\bmu^L}; {\bmu^R}} (\be)$ (\ref{mpcf}) is relied upon that developed in \cite{b1, b2} for
$G^{\,0}_{{\bmu^L}; {\bmu^R}} (\be)$ (\ref{mpcf2}). Indeed, differentiating (\ref{mpcf21}) over $\be$ and using the commutation relation
\begin{equation}
[ H_{\rm xx}, \si_{l_1}^{-} \si_{l_2}^{-} \dots \si_{l_N}^{-}] = \sum_{k=1}^N \si_{l_1}^{-} \dots \si_{l_{k-1}}^{-}\, [ H_{\rm xx}, \si_{l_k}^{-}]\, \si_{l_{k+1}}^{-} \dots\si_{l_N}^{-}  \label{comcom}
\end{equation}
together with $H_{\rm xx} \mid \Uparrow\rangle = 0$ and $\si^z_k \mid \Uparrow\rangle =\mid \Uparrow\rangle$,
one obtains the differential-difference equation
at fixed ${\bmu^L}$:
\begin{align}
\nonumber
\frac d{d \be}\,G_{{\bmu^L}; {\bmu^R}}(\be) & =\,h \Bigl(\frac{M}2 - N\Bigr)\, G_{{\bmu^L}; {\bmu^R}}(\be)
\\
\label{mpcf3} & +\,\frac 12 \sum_{k=1}^N \bigl(G_{{\bmu^L}; {\bmu^R}+{\bf e}_{k}} (\be)  + \,G_{{\bmu^L}; {\bmu^R}-{\bf e}_{k}} (\be)\bigr)\,,
\end{align}
where ${\bf e}_{k}$ is $N$-tuple
defined in (\ref{ratbe7272})
(and a similar equation for fixed ${\bmu^R}$). Equation (\ref{mpcf3}) is supplied with the initial condition $G_{{\bmu^L}; {\bmu^R}}(0) = \prod_{m=1}^N \dl_{\mu^L_m, \mu^R_m}$, as well as with the periodicity condition:
\begin{equation}
\label{ratbe7171}
G_{{\bmu^L}; {\bmu^R}} (\be) = G_{{\bmu^L}+ M {\bf e}_{k};\, {\bmu^R}} (\be) = G_{{\bmu^L};\, {\bmu^R}+ M {\bf e}_{k}} (\be)\,,\qquad \forall k\in\cE\,.
\end{equation}
Self-avoiding walks of vicious walkers are described by solution to (\ref{mpcf3}) provided that the non-intersection condition is imposed: $G_{{\bmu^L}; {\bmu^R}}(\be)=0$, if $\mu^R_k=\mu^R_p$ (or $\mu^L_k=\mu^L_p$) for any $1\leq k, p\leq N$.

The orthonormality relation is valid for the Schur functions (\ref{sch}):
\begin{equation}
\label{ratbe77}
\frac{1}{M^N}
\sum_{\{{\bphi}_N\}} |\CV (e^{i{\bphi}_N})|^2\,
S_{{\blad^L}}(e^{-i{\bphi}_N})\,
S_{{\blad^R}}(e^{i{\bphi}_N})
= \dl_{{\blad^L}, {\blad^R}} \,,
\end{equation}
where $\dl_{{\blad^L}, {\blad^R}}$ is unity for  coinciding ${\blad^L}$ and ${\blad^R}$ or zero otherwise.
The sum in (\ref{ratbe77}) is over $N$-tuples ${\bphi}_N = (\phi_{k_1}, \phi_{k_2}, \dots, \phi_{k_N})$,
where $\phi_n = \frac{2\pi}{M} \bigl(n-\frac{M}{2}\bigr)$ and $M\ge k_1>k_2> \cdots > k_N\ge 1$. Moreover, ${\CV} (e^{i {\bphi}_N})$ is defined by (\ref{spxx1}), and
$e^{\pm i {\bphi}} \equiv (e^{\pm i\phi_1}, e^{\pm i\phi_2}, \dots, e^{\pm i\phi_N})$. From (\ref{ratbe7272}) and (\ref{ratbe77}) one obtains, \cite{b1, b2}, the following

\vskip0.3cm \noindent
\noindent{\bf Statement~1:\,}
\textit{Solution to \eqref{mpcf3} respecting the initial condition $G_{{\bmu^L}; {\bmu^R}}(0) = \dl_{{\blad^L}, {\blad^R}}$, as well as the periodicity condition
\eqref{ratbe7171}, is of the form}:
\begin{equation}
G_{{\bmu^L}; {\bmu^R}} (\be)\,=\,
\displaystyle{\frac{1}{M^N}
\sum\limits_{\{{\bphi}_N\}}
e^{- \be E_N({\bphi}_N)}}|{\CV} (e^{i {\bphi}_N})|^2\,
S_{{\blad^L}}(e^{i {\bphi}_N})
S_{{\blad^R}}(e^{-i {\bphi}_N})\,,
\label{ratbe7}
\end{equation}
\textit{where
${\blad^{L, R}}={\bmu^{L, R}}- {\bdl}_N$, $E_N({\bphi}_N)$ is defined by \eqref{egen}, and the sum is the same as \eqref{ratbe77}}.

\vskip0.3cm \noindent
The solution to (\ref {mpcf3}) at $h=0$, which respects the non-intersection requirement, arises from (\ref {ratbe7}), and it is appropriate, \cite{forr2, b1, nest}, to provide it in the determinantal form:
\begin{equation}
G^0_{{\bmu^L}; {\bmu^R}}(\be)\,=\,\displaystyle{
\det \bigl(
{G}^0_{ {\mu_{n}^L}; {\mu_{k}^R}} (\be) \bigr)_{1\le
n, k \le N}}\,.
\label{ratbe6}
\end{equation}
Here ${G}^0_{j;\,m}(\be)$ is the solution to (\ref{mpcf3}) at $N=1$ and $h=0$:
\begin{equation}
\label{qanal277}
{G}^0_{j;\,m}(\be) \,=\, \displaystyle{
\frac{1}{M} \sum\limits_{n=1}^{M}
e^{\be \cos \phi_n}\,
e^{i \phi_n(m-j)}}\,,
\end{equation}
and $\phi_n = \frac{2\pi}{M}
\bigl(n-\frac{M}{2}\bigr)$.

Provided that $\frac{1}{M} \sum_{n=1}^{M}$ is replaced by $\frac{1}{2\pi}\int_{-\pi}^{\pi} dp$ at increasing $M$, the function ${G}^0_{j;\,m}(\be)$ (\ref{qanal277}) is reduced to the modified Bessel function of the first kind,
\begin{equation}
\label{ratbe7373}
{G}^0_{j;\,m}(\be) \,=\,I_{|j- m|}(\be)\,,
\end{equation}
where the power series is valid at $K+|j-m|= 0 \pmod{2}$:
\begin{equation}
I_{|j-m|}({\be})=\sum_{K\geq|m-j|}
\frac{({{\be}}/2)^K}{\bigl( \frac{K-|j-m|}2 \bigr)!\, \bigl(\frac{K+|j-m|}2 \bigr)!}\,.
\label{25}
\end{equation}
Assume that $\mathcal D^K_{s}$ is the $K^{\rm{th}}$ order differentiation with respect to $s$ at $s=0$. Application of  $\mathcal D^K_{s}$ to \eqref{25} gives the number ${|P_K(m\rightarrow j)|}$ of lattice paths consisting of $K$ steps between $m^{\rm th}$ and $j^{\rm th}$ sites on the infinite axis in terms of the binomial coefficient, \cite{b1}:
\begin{equation}
|P_K(m\rightarrow j)| =
\begin{pmatrix}
|m-j|+2L\\
L
\end{pmatrix}\,,\qquad
\begin{pmatrix}
K\\
L
\end{pmatrix} \equiv
\frac{K!}{L!\,(K-L)!}\,,
\label{26}
\end{equation}
where $L$ is one-half of the total number of turns:
$L\equiv(K-|m-j|)/2$.

\subsection{The random turns walks and the circulant matrix}
\label{ss341}

Acting by $\mathcal D^K_{\be/2}$ on $G_{{\bmu^L}; {\bmu^R}} (\be)$ (\ref{mpcf}) one obtains the average of $K^{\rm th}$ power of the total Hamiltonian:
\begin{equation}
\mathfrak{G} ({\bmu^L}; {\bmu^R}\,| K) \equiv \mathcal D^K_{\be/2} \,G_{{\bmu^L}; {\bmu^R}} (\be)
= \langle {\bmu^L} |\,(-2 H)^K  |\, {\bmu^R} \rangle
\label{mpcf44}\,.
\end{equation}
It follows from (\ref{mpcf44}) that
$\mathfrak{G} ({\bmu^L}; {\bmu^R}\,| 0) = \dl_{{\bmu^L}, {\bmu^R}}$
due to the orthogonality (\ref{tuc3}), where $\dl_{{\bmu^L}, {\bmu^R}}$ is unity for coinciding ${\bmu^L}$ and ${\bmu^R}$, or zero otherwise. With regard at (\ref{mpcf44}), let us represent the solution to
(\ref{mpcf3}) in the power series form:
\begin{equation}\label{new1}
G_{{\bmu^L}; {\bmu^R}} (\be)\,=\,
\sum_{K=0}^{\infty}
\frac{(\be/2)^K}{K!}\,
\mathfrak{G} ({\bmu^L}; {\bmu^R}\,| K)\,,
\end{equation}
where the coefficients $\mathfrak{G} ({\bmu^L}; {\bmu^R}\,| K)$ respect the equation which is due to substitution of (\ref{new1}) into (\ref{mpcf3}):
\begin{align}
\nonumber
\mathfrak{G} ({\bmu^L}; {\bmu^R}\,| K+1) & = h(M-2N)\, \mathfrak{G} ({\bmu^L}; {\bmu^R}\,| K) \\
& +\, \sum_{k=1}^{N} \bigl(\mathfrak{G} ({\bmu^L}; {\bmu^R}+{\bf e}_{k}\,| K) +
\mathfrak{G} ({\bmu^L}; {\bmu^R}-{\bf e}_{k}\,| K)\bigr)\,.
\label{mpcf6}
\end{align}
Equation (\ref{mpcf6}) is a difference version of (\ref{mpcf3}) of the type considered in \cite{forr2}. It is also supplied with the initial condition $\mathfrak{G} ({\bmu^L}; {\bmu^R}\,| 0) = \dl_{{\bmu^L}, {\bmu^R}}$, as well as with appropriate periodicity and non-intersection requirements.

Equation
(\ref{mpcf6}) at $h=0$
provides an ``isotropic'' version of a more general equation derived in \cite{forr2} for the random turns model with non-coincidence of the ``weights'' corresponding to left and right jumps of a randomly chosen walker. Furthermore, a comparison with \cite{forr2} demonstrates that not only jumps to neighboring sites are allowed, but there is an opportunity for all walkers to stay stationary since the spin chain is coupled to the homogeneous magnetic field $h$ (cf. Figure~\ref{fig:f1}).

Let us assume that $G^0_{{\bmu^L}; {\bmu^R}} (\be)$ (\ref{mpcf2}) is also given by the series analogous to (\ref{new1}), where the coefficients  $\mathfrak{G}^{\,0}({\bmu^L}; {\bmu^R}\,| K)$ are defined as follows:
\begin{equation}
\mathfrak{G}^{\,0}({\bmu^L}; {\bmu^R}\,| K)
\equiv \mathcal D^K_{\be/2}\, G^{\,0}_{{\bmu^L}; {\bmu^R}} (\be) = \langle {\bmu^L} |\,(-2 H_{\rm xx})^K  |\, {\bmu^R} \rangle \,.\label{qanal13}
\end{equation}
The average  $\mathfrak{G}^{\,0}({\bmu^L}; {\bmu^R}\,| K)$ respects (\ref{mpcf6}) at $h=0$ since
$G^0_{{\bmu^L}; {\bmu^R}} (\be)$ is described by (\ref{mpcf3}) at $h=0$. Expanding the exponential in (\ref{mpcf21}) and taking (\ref{new1}) into account, one obtains the identity:
\begin{equation}
\mathfrak{G} ({\bmu^L}; {\bmu^R}\,| K)\,
 =\,\sum_{i=0}^{K}
\begin{pmatrix}
K\\
i
\end{pmatrix}
\bigl(h(M-2N) \bigr)^i\,
\mathfrak{G}^{\,0}({\bmu^L}; {\bmu^R}\,| K-i)\,,
\label{mpcf4}
\end{equation}
where $\begin{pmatrix}
K\\ i
\end{pmatrix}$ is the binomial coefficient (\ref{26}). Right-hand side of (\ref{mpcf4}) is reduced at $h=0$ to $\mathfrak{G}^{\,0} ({\bmu^L}; {\bmu^R}\,| K)$ since only $i=0$ contributes.

The circulant matrix
${\bold\Delta}$ (\ref{cor:lin4}) leads to the $N=1$ solution of (\ref{mpcf6}) at $h=0$:
\begin{equation}
\mathfrak{G}^0 (j, m | K) =  \langle \Uparrow \mid \si_j^{+} (-2 H_{\rm xx})^K \si_m^{-} \mid
\Uparrow \rangle = \bigl({\bold \Delta}^K \bigr)_{j m}\,,
\label{avv}
\end{equation}
where
$\bigl({\bold \Delta}^K \bigr)_{j m}$ is the entry of $K^{\rm th}$ power of
${\bold\Delta}$, which fulfils
\begin{equation}
\label{dmcf}
\bigl({\bold \Delta}^{K+1} \bigr)_{j m}
= \bigl({\bold \Delta}^{K} \bigr)_{j, m+1} +
\bigl({\bold \Delta}^{K} \bigr)_{j, m-1}\,.
\end{equation}
The initial condition is respected since $\mathfrak{G}^0 (j, m | 0)$ is the Kronecker symbol $\delta_{j m}$. The periodicity requirement is also consistent with the circulant matrix (\ref{cor:lin4}).

Position of the walker on the chain is labelled by the spin ``down''
state, while the empty sites correspond to spin ``up'' states. Let $|P^0_K(j \rightarrow m)|$ to denote the number of $K$-step paths of a single walker between $j^{\rm {th}}$ and $m^{\rm {th}}$ sites ($h=0$). Evaluation of (\ref{qanal13}) corresponding to $N=1$ results in $|P^0_K(j \rightarrow m)|= \bigl({\bold\Delta}^K \bigr)_{j m}$ in agreement with (\ref{avv}).

Let us turn to the lattice paths made by $N$ vicious walkers with initial and final positions arranged as the strict partitions ${\bmu^L}$ and ${\bmu^R}$, respectively, and let
$|P^0_K ({\bmu^L} \rightarrow\, {\bmu^R})|$ be the number of sets of paths characterized by the total number of steps $K$. We formulate the following

\vskip0.3cm \noindent
\noindent{\bf Proposition~3:\,}
\textit{The number of sets of self-avoiding lattice paths of $N$ vicious walkers with the total number of steps $K$ is equal to the amplitude $\mathfrak{G}^{\,0} ({\bmu^L}; {\bmu^R}\,| K)$ solving \eqref{mpcf6} at $h=0$}:
\begin{align}
\nonumber
|P^0_K ({\bmu^L} \rightarrow\,{\bmu^R})| & =\,
\mathfrak{G}^{\,0}({\bmu^L}; {\bmu^R} | K) \\
\label{qanal272}
& =\,\sum_{|{\bf n}|=K} P({\bf n})\,
\det\bigl(({\bold\Delta}^{n_j} )_{\mu^L_{i}; \mu^R_j}\bigr)_{1
\le i, j \le N}\,,
\end{align}
\textit{where ${\bf n}=(n_1, n_2, \ldots, n_N)$, $|{\bf n}|\equiv n_1+n_2+ \ldots + n_N$, $P({\bf n})$ is the multinomial coefficient},
\begin{equation}
\label{qanal271}
P({\bf n}) \equiv
\frac{(n_1+n_2+ \ldots+n_N)!}{n_1!\, n_2!\, \dots n_N!}\,,
\end{equation}
\textit{the entry $({\bold \Delta}^{n})_{j m}$ is defined by \eqref{avv}, and $({\bold \Delta}^{0})_{j m} = {\delta}_{j m}$}.

\vskip0.3cm \noindent
\noindent{\bf Proof:\,} Equation (\ref{qanal272}) is reduced at $K=0$ to the orthogonality (\ref{tuc3}) and conjectured at arbitrary $K$ due to validity of (\ref{qanal13})
(see {\sf Appendix~I}). Here we shall verify that (\ref{qanal272}) indeed respects (\ref{mpcf6}) ($h=0$) as the generalization of (\ref{dmcf}) at $N>1$.

Induction with respect of $N$ enables to prove Eq.~(\ref{qanal272}) provided that the base case $N=1$
is given by (\ref{avv})
and (\ref{dmcf}). The induction step is to assume that (\ref{qanal272}) fulfills
(\ref{mpcf6}) ($h=0$), where the partitions $\bmu^L$ and $\bmu^R$ are of the length $N-1$
so that the minors in
(\ref{qanal272}) are of the size $(N-1)\times(N-1)$.

The proof is based on the identity for $\mathfrak{G}^{\,0} ({\bmu^L}; {\bmu^R}\,| K+1)$, where $\bmu^L$ and $\bmu^R$ are of the length $N$:
\begin{equation}
\mathfrak{G}^{\,0} ({\bmu^L}; {\bmu^R}\,| K+1) \, =\,\sum_{p=0}^{K+1} \begin{pmatrix}
K+1\\
p
\end{pmatrix} \sum_{m=1}^{N} (-1)^{N+m}\,\mathfrak{G}(K+1-p,p,m) \,.
\label{new2}
\end{equation}
The shortening notation is introduced in \eqref{new2}:
\begin{equation}
\mathfrak{G} (K,P,m)
\equiv \mathfrak{G}^{\,0} ({{\overset m \bmu}^{_{\,L}}_{N-1} }; {\bmu^R_{N-1}}\,| K)\,\mathfrak{G}^{\,0} ({\mu^L_m}; {\mu^R_N}\,| P)\,,
\label{new21}
\end{equation}
where ${{\overset {m} \bmu}^{_{\,L}}_{N-1}}\equiv (\mu^L_1, \mu^L_2,
\dots, \mu^L_{m-1}, \mu^L_{m+1}, \mu^L_N)$, $\bmu^R_{N-1} = (\mu^R_1, \mu^R_2, \ldots, \mu^R_{N-1})$, and $K$, $P$, $m$ are non-negative integers. The identity (\ref{new2}), (\ref{new21}) is due to expanding the determinant in (\ref{qanal272}) along $N^{\rm th}$ column.

The identity (\ref{new2}) is used
in right-hand side of equation (\ref{mpcf6}) (after $K+1 \mapsto K$), and this gives:
\begin{align}
\nonumber
\sum_{k=1}^{N}\mathfrak{G}^{\,0} ({\bmu^L}; {\bmu^R}\pm {\bf e}_k\,| K) & =\,
\sum_{p=0}^{K} \begin{pmatrix}
K\\
p
\end{pmatrix} \sum_{m=1}^{N} (-1)^{N+m} \\
\nonumber
& \times \mathfrak{G}^{\,0} ({\mu^L_m}; {\mu^R_N}\,| p) \sum_{k=1}^{N-1} \mathfrak{G}^{\,0} ({{\overset m \bmu}^{_{\,L}}_{N-1}}; {\bmu^R_{N-1}}\pm {\bf e}_k\,| K-p) \\
\nonumber
& +\,\sum_{p=0}^{K} \begin{pmatrix}
K\\
p
\end{pmatrix} \sum_{m=1}^{N} (-1)^{N+m} \\
\label{new7}
& \times \mathfrak{G}^{\,0} ({\mu^L_m}; {\mu^R_N}\pm 1\,| p)\,\mathfrak{G}^{\,0} ({{\overset m \bmu}^{_{\,L}}_{N-1}}; {\bmu^R_{N-1}}\,| K-p)
\,.
\end{align}

In turn, the series (\ref{new2}) itself is represented as
\begin{align}
\label{new3}
\mathfrak{G}^{\,0} ({\bmu^L}; {\bmu^R}\,| K+1) & =\,\sum_{m=1}^{N} (-1)^{N+m}
\mathfrak{G} (K+1, 0, m)
\\
&+\,\sum_{p=1}^{K} \begin{pmatrix}
K+1\\
p
\end{pmatrix} \sum_{m=1}^{N} (-1)^{N+m} \mathfrak{G} (K+1-p,p,m)
\label{new4}
\\
\label{new5}
& +\,\sum_{m=1}^{N} (-1)^{N+m}
\mathfrak{G} (0, K+1, m)\,,
\end{align}
where the notation \eqref{new21} is used. Further, the Pascal relation
\begin{equation}
\label{new6}
\begin{pmatrix}
K+1\\
p
\end{pmatrix}\,=\,\begin{pmatrix}
K\\
p-1
\end{pmatrix}\,+\,
\begin{pmatrix}
K\\
p
\end{pmatrix}
\end{equation}
is used in the line (\ref{new4}). The representation (\ref{new3}), (\ref{new4}), (\ref{new5})
is compared with two sums in right-hand side of (\ref{new7}) so that (\ref{new3}) is matched to $p=0$ in the first sum, and (\ref{new5}) is matched to $p=K$
in the second sum. The base case is applied to $\mathfrak{G}^{\,0} ({\mu^L_m}; {\mu^R_N}\,| p)$ in the contribution corresponding to the first term in (\ref{new6}), whereas the induction assumption
is applied to $\mathfrak{G}^{\,0} ({{\overset m \bmu}^{_{\,L}}_{N-1}}; {\bmu^R_{N-1}}\,| K+1-p)$ in the contribution corresponding to the second term in (\ref{new6}).
The coincidence of $\mathfrak{G}^{\,0} ({\bmu^L}; {\bmu^R}\,| K+1)$ with the sum of two identities (\ref{new7}) is thus established.

The determinantal expression (\ref{qanal272}) ensures validity of the non-intersection requirement and provides the number $|P^0_K ({\bmu^L} \rightarrow\,{\bmu^R})|$ of $K$-step sets of paths traced
by $N$ vicious walkers. $\Box$

Right-hand side of (\ref{mpcf4}) is re-arranged as the polynomial of two variables, $h(M-N)$ and $-hN$:
\begin{align}
\nonumber
P_K ({\bmu_N^L}, {\bmu_N^R}) & \equiv\,\sum_{p_1+p_2+p_3=K}
P (p_1, p_2, p_3) \\
\label{mpcf5}
& \times\,|P^0_{p_3} ({\bmu_N^L} \rightarrow\,{\bmu_N^R})|\,
\bigl(h(M-N)\bigr)^{p_1}\,
(-hN)^{p_2}\,,
\end{align}
where the coefficient $P (p_1, p_2, p_3)$ is defined by (\ref{qanal271}). The coefficients
$|P_{p_3}^0 ({\bmu^L} \rightarrow\, {\bmu^R})|$ enumerate, due to
\textsf{Proposition~3},  $p_3$-step sets of paths of $N$ walkers. Recall that either a single walker chosen randomly jumps to one of closest sites with equal probabilities or all walkers are staying stationary. Therefore $P_K ({\bmu_N^L}, {\bmu_N^R})$ (\ref{mpcf5}) corresponds to a superposition of sets of $(K-p_1)$-step paths at each fixed $0\le p_1\le K$. The product
$P (p_1, p_2, p_3) |P^0_{p_3} ({\bmu_N^L} \rightarrow\, {\bmu_N^R})|$ gives the number of sets of $N$ paths such that $p_3$ times one walker jumps and $p_2$ times all walkers are staying stationary ($p_2+p_3 = K - p_1$). A typical configuration of $N=6$ paths for $p_1=0$, $p_2=1$, and $p_3=K-1$  is shown in Fig.~\ref{fig:f1} ($K=13$) where dashed lines imply that walkers are staying. As far as $|P_{p_3}^0 ({\bmu^L} \rightarrow\, {\bmu^R})|$ (\ref{qanal272}) is concerned, the configuration in Fig.~\ref{fig:f1} corresponds to $n_1=0$, $n_2=1$, $n_3=3$, $n_4=1$, $n_5=4$, $n_6=3$.

\subsection{Generalized
Ramus's identity}
\label{s342}

The present section is devoted to a relationship between the powers of the circulant matrix ${\bold\Delta}$ \eqref{cor:lin4} and the binomial coefficients expressing the numbers of lattice paths \eqref{26}.

Calculation of the entries of integer positive powers of circulant matrices attracts attention \cite{rim1, rim2, feng}. For instance, the entries
$\bigl({\bold \Delta}^K \bigr)_{j m}$ at $K$ arbitrary are obtained in \cite{rim1, rim2} for ${\bold \Delta}$ of even order in terms of the Chebyshev polynomials. In the present paper expression of $\bigl({\bold\Delta}^K \bigr)_{j m}$ by means of \textit{Ramus's identity} \cite{ram} is used (cf. \cite{ram1, ram2}). The latter provides the entries in terms of the binomial coefficients thus stressing the connection with enumeration of the lattice walks (cf.~(\ref{26})).

The vanishing $\bigl({\bold\Delta}^K \bigr)_{j m}=0$ occurs for the circulant matrix (\ref{cor:lin4}) in the case $K-|j-m| = 1 ({\rm mod}\,2)$. In the case $K - |j-m| =0 ({\rm mod}\,2)$, the Ramus's identity (see \textsf{Appendix II}) allows us to formulate

\vskip0.3cm \noindent
\noindent{\bf Proposition~4:\,}
\textit{The row-column indices $j, m$ of $M\times M$ matrix respect $|j-m|\le M-1$. Let us assume that $L\equiv \frac{K-|j-m| \pm p M}{2}$ is chosen so that $0\le L\le \frac{M}{2}$ and $p\in\BN$. Then},
\begin{equation}
\label{qanal276}
\bigl({\bold\Delta}^K \bigr)_{j m}\, =\,
\begin{pmatrix} K \\ L {\bar\dl}_{L, \frac{M}{2}}
\end{pmatrix}_{{M}/{2}}\,,
\end{equation}
\textit{where ${\bar\dl}_{L, \frac{M}{2}}\equiv 1-{\dl}_{L, \frac{M}{2}}$, and
the notation for the lacunary sum of binomial coefficients is used}, \cite{ram3}:
\begin{equation}
\begin{pmatrix} K \\ L \end{pmatrix}_{{M}/{2}} \equiv
\begin{pmatrix} K \\ L \end{pmatrix} +
\begin{pmatrix} K \\ L +\frac{M}{2}\cdot 1 \end{pmatrix} +
\begin{pmatrix} K \\  L +\frac{M}{2}\cdot 2 \end{pmatrix} + \ldots \,.
\label{qanal27611}
\end{equation}

\vskip0.3cm \noindent
\noindent{\bf Proof:\,}
The transition element (\ref{qanal13}) arising from (\ref{qanal277}) takes the form
(recall that $M$ is even):
\begin{align}
\mathfrak{G}^{\,0}(j; m | K)
= \frac{2^{K+1}}{M}
\sum_{l=0}^{\frac{M}{2}-1}
\cos^K \Bigl(\frac{2\pi l}{M}\Bigr)
\cos\Bigl(\frac{2\pi l |m-j|}{M}\Bigr)
\nonumber \\
+\,\bigl((-1)^{K+|m-j|} - 1\bigr)
\frac{2^{K}}{M}\,.
\label{qanal278}
\end{align}
The Ramus's identity ({\rm AII.1}) allows us to re-express the series in (\ref{qanal278}) provided that $n$ and $n-2t$ are replaced by $K$ and $K-2L$, respectively. As the result, the validity of (\ref{qanal276}) is verified for $K - |m-j|=0 ({\rm mod}\,2)$ at $L\ne \frac{M}{2}$. As it is clear from (\ref{qanal278}) and ({\rm AII.1}), the equivalence of the cases
$L = \frac{M}{2}$ and $L=0$ confirms the validity of (\ref{qanal276}). A trigonometric transformation of (\ref{qanal278}) allows us also to demonstrate that $\bigl({\bold\Delta}^K \bigr)_{j m}=0$ at $K-|j-m| = 1 ({\rm mod}\,2)$.\footnote{See \textsf{Appendix III} for illustrative examples.} $\Box$

{\sf Proposition~4} demonstrates that Ramus's identity allows one to express the entries of ${\bold\Delta}^K$ as the lacunary sums of the binomial coefficients. On another hand, ${\bold\Delta}^K$ respects (\ref{dmcf}) which is the particular case of
(\ref{mpcf6}) at $h=0$. Therefore it looks appropriate to relate (\ref{mpcf6}) at arbitrary $N$
with appropriate generalized Ramus's identities. Regarding at (\ref{ratbe7}) and (\ref{qanal272}), we formulate

\vskip0.3cm \noindent {\bf Proposition~5\,} \textsf{(generalized Ramus's identity):\,}
\textit{The following identity is valid}:
\begin{align}
\sum_{|{\bf n}|=K} P({\bf n})\,
{\bold\Delta}^{\bf n}_{{\bmu^L}; {\bmu^R}} &=\,
\displaystyle{\frac{2^{K+N}}{ M^N}} \sum \limits_{{\bf{l}} \in {\sf P}^N}
\left(\sum\limits_{k=1}^N \cos \Bigl(\frac{2\pi}{M} {l_k}\Bigr) \right)^K \nonumber \\
&\times\,\prod_{s=1}^{N}
\cos\Bigl(\frac{2\pi}{M} {l_s}
({\mu_{s}^L}-{\mu_s^R}) \Bigr) \,,
\label{ratbeco1}
\end{align}
\textit{where}
\begin{equation}
\label{qanal27}
{\bold\Delta}^{\bf n}_{{\bmu^L}; {\bmu^R}} \equiv \prod_{j=1}^{N}  ({\bold\Delta}^{n_j} )_{\mu^L_{j}; \mu^R_j} \,,
\end{equation}
\textit{$({\bold \Delta}^{n})_{j m}$ is defined by \eqref{qanal276},  \eqref{qanal27611},
$({\bold \Delta}^{0})_{j m} = {\delta}_{j m}$, whereas ${n_j}$ and $|\mu^L_{j}- \mu^R_j|$ are of the same parity. Summation is over $N$-tuples ${\bf l} \equiv ({l_1},
{l_2}, \ldots, {l_N})$, $l_k \in {\sf P}\equiv \{0, 1, \ldots, \frac{M}{2}-1\}$}.

\vskip0.3cm \noindent
\noindent{\bf Proof:\,}
Equation (\ref{ratbeco1}) is reduced at $N=1$ to Ramus's identity ({\rm AII.1}) although is directly verified at $N=2$. Mathematical induction with respect of $N$ is straightforward and relies upon the fact that left-hand side of (\ref{ratbeco1}) is represented at any $1\le m\le N$:
\begin{equation}
\nonumber
\sum_{p=0}^{K} \begin{pmatrix}
K \\
p \end{pmatrix}
{\Delta}^{p}_{{\mu^L_{m}}; {\mu^R_{m}}} \sum_{|{ \overset m {\bf n}}_N|=K-p} P({ \overset m {\bf n}}_N)\,
{\bold\Delta}^{{ \overset m {\bf n}}_N}_{
{{\overset m \bmu}^{_{\,L}}_{N}}; {{\overset m \bmu}^{_{\,R}}_{N}} }\,,
\end{equation}
where ${{\overset m \bmu}^{_{\,L}}_{N}}$, ${{\overset m \bmu}^{_{\,R}}_{N}}$ are defined in (\ref{new2}), and ${ \overset m {\bf n}}_N \equiv (n_1, n_2, \ldots, n_{m-1}, n_{m+1}, \ldots, n_N)$.
$\Box$

\vskip0.3cm \noindent {\bf Corollary:\,}

\noindent\textit{$\bullet\,\,$
Determinantal generalization of \eqref{ratbeco1} reads}:
\begin{align}
\sum_{|{\bf n}|=K} P({\bf n})\,
\det\bigl(({\bold\Delta}^{n_j} )_{\mu^L_{i}; \mu^R_j}\bigr)_{1
\le i, j \le N} &=\,
\displaystyle{\frac{1}{M^N}}
\sum\limits_{\{{\bphi}_N\}}
\Bigr(2\sum\limits_{m=1}^N \cos {\phi}_{m}\Bigl)^K \nonumber \\
&\times\,|{\CV} (e^{i {\bphi}_N})|^2\,
S_{{\blad^L}}(e^{i {\bphi}_N})
S_{{\blad^R}}(e^{-i {\bphi}_N})\,,
\label{ratbeco}
\end{align}
\textit{where the entries $({\bold\Delta}^{n_j} )_{\mu^L_{i}; \mu^R_j}$, $1
\le i, j \le N$, are given by} (\ref{qanal276}).

\noindent\textit{$\bullet\,\,$
The Schur functions are equal to unity, $S_{{\blad^L}}(e^{i {\bphi}_N})=S_{{\blad^R}}(e^{-i {\bphi}_N})=1$, provided that ${\bmu^L} = {\bmu^R} = \bdl_N$, where $\bdl_N$ is defined by \eqref{strict}. Then, Eq.~\eqref{ratbeco}
gives the number of self-avoiding trajectories of $N$ random turns walkers initially located at $\bdl_N$ and returning to their initial positions after $K$ steps over long enough chain} ($M\gg 1$):
\begin{align}
\label{ratbe66}
&\sum_{|{\bf n}|=K} P({\bf n})\,
\det\left(\begin{pmatrix}{n_j} \\ \frac{{n_j}+j-i}2
\end{pmatrix}\right)_{1
\le i, j \le N}\,=\,2^K
{\cal J} (K, N)\,, \\
\label{ratbe661}
{\cal J} (K, N) &\equiv\,
\displaystyle{\frac{1}{N!}}\,
\int\limits_{-\pi}^{\pi}
\int\limits_{-\pi}^{\pi}\dots
\int\limits_{-\pi}^{\pi}
\Bigr(\sum\limits_{m=1}^N \cos {\phi}_{m}\Bigl)^K |{\CV} (e^{i {\bphi}_N})|^2\,\frac{d\phi_1 d\phi_2\dots
d\phi_N}{(2\pi)^N}\,,
\end{align}
\textit{where zero values are assigned to the entries of the matrix in \eqref{ratbe66} provided that $n_j$ and $|i-j|$ are of opposite parity. Besides, when $n_j$ vanishes at some $j$, the entry of the matrix
is Kronecker symbol $\dl_{i j}$}.

\vskip0.3cm The integral ${\cal J} (K, N)$ (\ref{ratbe661}) is zero for $K$ odd (the same is true for the series in left-hand side of (\ref{ratbe66})), whereas ${\cal J} (K, N)$ is related at $K$ even with the number of
random permutations of $\{1, \ldots, \frac{K}{2}\}$ with at most $N$ increasing subsequences
\cite{forr}, as well as with the distribution of the length of the longest increasing subsequence of random permutations of $\{1, \ldots, \frac{K}{2}\}$ \cite{joh, joh1}. The problem of the longest increasing subsequence of random permutations is related to the random unitary matrices \cite{rain}, whereas more on connection of the longest increasing subsequence with various areas of mathematics can be found in \cite{dan}.

\subsection{Transition amplitude as the generating function of random turns walks}
\label{s333}

With regard at {\sf Proposition~3}, let us turn to the representation (\ref{new1}). Provided that the numbers $|P^0_K ({\bmu^L_N} \rightarrow {\bmu^R_N})|$ (\ref{qanal272}) taken at $M\to\infty$ are considered as coefficients of the power series in $\beta$, one meets the following

\vskip0.3cm \noindent {\bf Proposition~6:\,}
\textit{The determinantal representation
\begin{equation}
\label{ratbe6621}
\sum_{K=0}^{\infty} \frac{(\be/2)^K}{K!}\,
|P^0_K ({\bmu^L_N}\,\rightarrow\, {\bmu^R_N})|\,
=\,\det\bigl(I_{|{ {\mu_{i}^L}- {\mu_{j}^R}}|}(\be)\bigr)_{1
\le i, j \le N}\,,
\end{equation}
where $I_{|{ {\mu_{i}^L}- {\mu_{j}^R}}|}(\be)$ is the modified Bessel function of the first kind, is valid for the  power series provided that its coefficients are given by \eqref{qanal272} with the entries \eqref{qanal276} taken in the form} (\ref{26}).

\vskip0.3cm \noindent
\noindent{\bf Proof:\,}
As the base case, Eq.~(\ref{ratbe6621}) is verified at $N=2$ with the usage of (\ref{25}), (\ref{26}) in its right-hand side. Assume that (\ref{ratbe6621}) is valid for $(N-1)^{\rm th}$ order. To express the induction step, we re-express left-hand side of (\ref{ratbe6621}):
\begin{equation}\label{ratbe66211}
\sum_{p=0}^{\infty}
\sum_{K\ge p}^{\infty}\frac{(\be/2)^K}{K!}\,
|P^0_{K} ({\bmu^L_N}\,\rightarrow\, {\bmu^R_N})|_{n_{_N}\equiv\, p}\,,
\end{equation}
where expansion of the determinant in $|P^0_{K-p} ({\bmu^L_N}\,\rightarrow\, {\bmu^R_N})|_{n_{_N}\equiv\, p}$
along $N^{\rm th}$ column takes the form:
\begin{align}
\nonumber
|P^0_{K} ({\bmu^L_N}\,\rightarrow\, {\bmu^R_N})|_{n_{_N}\equiv\, p} & =\,\begin{pmatrix}
K\\
p
\end{pmatrix} \sum_{m=1}^{N} (-1)^{N+m} \\
\label{new8}
& \times |P^0_{K-p} ({{\overset m \bmu}^{\,L}_{N-1}}\,\rightarrow\, {\bmu^R_{N-1}})|\,
|P^0_{p} ({\mu^L_m}\,\rightarrow\, {\mu^R_N})|\,.
\end{align}
Using the base case together with the induction assumption
to express the infinite series (\ref{ratbe66211}), one obtains the corresponding expansion of the determinant in right-hand side of (\ref{ratbe6621}) along
$N^{\rm th}$ column. $\Box$

Equation (\ref{ratbe6621}) generalizes the case of $N=1$
corresponding to Eqs.~(\ref{25}), (\ref{26}). The Bessel function of the first kind as the generating function of sets of paths between two sites of infinite chain has been discussed in \cite{b1}. Equation (\ref{ratbe6621}) reads:

\noindent $\bullet\,$ \textit{The determinant of the Bessel functions is the generating function of the numbers of $K$-step sets of paths $|P^0_K ({\bmu^L_N} \rightarrow {\bmu^R_N})|$}.

According to {\sf Proposition~6}, one gets in the particular case ${\bmu^L_N} = {\bmu^R_N} = \bdl_N$:
\begin{align}
\label{ratbe662}
\sum_{K=0}^{\infty}
\frac{(\be/2)^K}{K!}\,
|P^0_K ({\bdl}_N \rightarrow\,{\bdl}_N)|\,
=\, z(2/\be, N)\,,\\
\label{ratbe663}
z(2/\be, N) \, \equiv\, \det\bigl(I_{|i-j|}(\be)\bigr)_{1
\le i, j \le N}\,,
\end{align}
where $z(2/\be, N)$ (\ref{ratbe663}) coincides with  the correlation function $G^0_{{\bdl}_N; {\bdl}_N}(\be)$ (\ref{ratbe6}) at large enough $M$.
In other words,
$z(2/\be, N)$ coincides with the Gross-Witten partition function, which demonstrates a third order phase transition at $N\to \infty$, \cite{gross}.
Connection between the $XX$ spin chain and the low-energy QCD, as well as a possibility of a third order phase transition in the spin chain, are discussed in \cite{tier, zah}.

\section{The averages over Bethe state-vectors and nests of lattice paths}
\label{ss342}

Let us begin with the calculation of the normalized average of the generating exponential over the Bethe state-vectors given by {\sf Definition~1}:
\begin{equation}
\langle e^{\cal Q}\rangle_N \equiv \frac{\langle \Psi (e^{i{\bth}_N/2}) \!\mid
e^{\cal Q} \mid\! \Psi (e^{i{\bth}_N/2}) \rangle}{\CN^2 (e^{i{\bth}_N/2})}\,, \label{ratbe0}
\end{equation}
where ${\cal Q}$ is given by (\ref{cor:lin05}), and $N$-tuple $e^{{i\bth}_N/2} = (e^{i\ta_{1}/2}, e^{i\ta_{2}/2}, \ldots , e^{i\ta_{N}/2})$ is to express
the substitute $v_j = u_j\equiv
e^{i\ta_{j}/2}$ ($1\le j\le N$).
Using {\sf Proposition~1} and the Bethe solution (\ref{besol}), we express $\langle e^{\cal Q}\rangle_N$ (\ref{ratbe0}):
\begin{equation}
\langle e^{\cal Q}\rangle_N =
\det(e^{\widehat\al})\,,\qquad
e^{\widehat\al} \equiv \Bigl(
\displaystyle{ \frac{1}{M}\sum\limits_{n=1}^{M} e^{\al_n + i n (\theta_i-\theta_j)}} \Bigr)_{1\le i,j \le N}\,.
\label{cor:lin55311}
\end{equation}

The generating exponential $e^{\cal Q}$ under the conventional
specialization of ${\bf a}_M$ is replaced by $e^{\al Q(m)}$, and the average $\langle e^{\al Q(m)}\rangle_N$ (\ref{cor:lin55311}) is known as the generating function of mean values of third components of spins \cite{vk1, ess, ml3, col, col1}:
\begin{align}
\nonumber
&\frac{\langle \Psi (e^{i{\bth}_N/2}) \!\mid e^{\al Q(m)}
\mid\! \Psi (e^{i{\bth}_N/2}) \rangle}{\CN^2 (e^{i{\bth}_N/2})}\,= \\
& =\,\det
\begin{pmatrix} \displaystyle{
\Bigl(1+(e^\al-1) \frac{m}{M} \Bigr) \dl_{ij}} + (e^\al-1) (1-\dl_{ij})\,Q_{\ta_i, \ta_j}(m) \end{pmatrix}_{1\le i,j
\le N}\,,
\label{cor:lin556}
\end{align}
where
\begin{equation}
\nonumber
Q_{\ta_i, \ta_j}(m)\,\equiv\,\frac{1}{M} \frac{\sin
\frac{m(\ta_i-\ta_j)}{2} }{\sin \frac{\ta_i-\ta_j}{2}}\,.
\end{equation}
Invariance of the determinant (\ref{cor:lin556}) under conjugation of the matrix by the diagonal matrix $e^{i\frac{m}{2}\h\ta}$ (where $\h\ta\equiv \diag\{ \ta_i\}$) is accounted for.

Let us obtain the Boltzmann-weighted average of $e^{\cQ}$ with respect of the Bethe state-vectors
characterized by {\sf Definition~1}.
A determinantal expression for the corresponding off-shell average is calculated by insertion of the
decomposition of unity (\ref{field7}):
\begin{align}
\nonumber
& \langle \Psi({\bf v}_N) |
e^{\cQ}\,e^{-\be H} | \Psi({\textbf u}_N)\rangle\,=\,\sum \limits_{\{{\bphi}_N\}}
\frac{e^{-\be E_N ({\bphi}_N)}}{\CN^{2} (e^{i{\bphi}_N/2})} \\
& \times\,\langle \Psi({\bf v}_N) | e^{\cQ} | \Psi(e^{i{\bphi}_N/2})\rangle \,\frac{\det T_M(e^{-i{\bphi}_N}, {\textbf u}^2_N)}{
{\CV} (e^{-i{\bphi}_N}) {\CV} ({\textbf u}^2_N)}\,,
\label{cor:lin661}
\end{align}
where $E_N ({\bphi}_N)$ is given by (\ref{egen}), and (\ref{spxx3}), (\ref{cauchy}),  (\ref{cauchy1}), (\ref{spxx}), and (\ref{ratbe91}) are accounted for.
Taking into account {\sf Proposition~1} to express $\langle \Psi({\bf v}_N) | e^{\cQ} | \Psi(e^{i{\bphi}_N/2})\rangle$, one obtains:
\begin{align}
\nonumber
& \langle \Psi({\bf v}_N) |
e^{\cQ}\, e^{-\be H} | \Psi({\textbf u}_N)\rangle\,=\, \\
& =\,\displaystyle{ \frac{e^{\be h M/2}}{{\CV}({\textbf
u}_N^2){\CV}({\textbf v}_N^{-2})} \det
\left(\sum\limits_{k, l=1}^{M}
e^{{\al}_{_k}}
{G}_{k;\,l}(\be)\, \frac{u_i^{2 l}}{v_j^{2k}} \right)_{1\le i,j \le N}}\,,
\label{cor:lin663}
\end{align}
where
\begin{equation}
\label{qanal27777}
{G}_{k;\,l} (\be) \,\equiv\, \displaystyle{
\frac{1}{M} \sum\limits_{p\in {\sf S}^\pm}
e^{-\be\ep (p)}\,
e^{i p (l-k)}}\,.
\end{equation}
Summation in (\ref{qanal27777}) is over either of two sets ${\sf S}^\pm\ni p$ specified by
$\cos Mp = \mp 1$:
\begin{equation}
\begin{array}{l}
{\sf S}^{+} =\bigl\{-\pi+ \frac{2\pi}{M}(n-\frac12) \bigr\}_{n \in {\cal E}}\,, \\[0.2cm]
{\sf S}^{-} =\bigl\{-\pi + \frac{2\pi}{M} n \bigr\}_{n \in {\cal E}}\,,
\end{array}
\label{cor:linn23}
\end{equation}
and the choice of ${\sf S}^+$ or ${\sf S}^-$ is due to evenness or oddness of $N$ in (\ref{cor:lin663}).
Equation (\ref{cor:lin663}) on solution to the Bethe equations leads to the normalized average:
\begin{align}
\nonumber
\langle e^{\cQ}\, e^{-\be H} \rangle_N & \equiv \frac{\langle \Psi (e^{i{\bth}_N/2}) | e^{\cQ}\, e^{-\be H} | \Psi (e^{i{\bth}_N/2}) \rangle}{\CN^2 (e^{i{\bth}_N/2})} \\
 & =\,e^{\be h M /2}\,\det
\bigl( e^{- \be {\h\ep}} {e^{\h\al}}\bigr)\,,
\label{cor:lin665}
\end{align}
where $e^{\h\al}$ is defined by (\ref{cor:lin55311}), and the diagonal matrix
${\h\ep}$ consists of
$\ep(\ta_j)$ (\ref{egen}):
\begin{equation}
\label{mpcf1}
{\h\ep}\equiv \underset{1\le j\le N}{\diag} \{\ep(\ta_j)\}\,.
\end{equation}


The commutation relation (\ref{cor:lin551}) together with Eqs.~(\ref{mpcf}), (\ref{ratbe7}) allows us to obtain $\langle e^{\cal Q}\, e^{-\be H}\rangle_N$
in the integral form at $M\gg 1$:
\begin{align}
\nonumber
\langle e^{\cal Q}\, e^{-\be H}\rangle_N
\,\simeq\,
\frac{e^{\be h (\frac M2 -N)}}{{\CN}^2 (e^{i{\bth}_N/2}) N!}\,\int\limits_{I_N}
{\cal P}_{\CK}(e^{-i{\bf p}_N}, e^{i{\bth}_N}, {\bf 0}_M)
\\
\times \,{\cal P}_{\CK}
(e^{-i{\bth}_N}, e^{i{\bf p}_N}, {\bf a}_M)\,| \CV (e^{i{\bf p}_N}) |^2\,
e^{{\be}\sum_{l=1}^N
\cos{p}_l}
\frac{d^Np}{(2\pi)^N}\,,
\label{cor:lin667}
\end{align}
where ${\bf p}_N\equiv (p_1,
p_2, \dots ,
p_N)$, $d^Np = d p_1 d p_2\cdots
d p_N$. The integration in (\ref{cor:lin667}) is over $N$-fold product $I_N\equiv \overset{{_N}}{\times}{\sf S}$
of the segment ${\sf S}\equiv [-\pi, \pi ]$. With regard at  (\ref{ratbe6}) and (\ref{ratbe7373}), the representation (\ref{cor:lin667}) takes the following equivalent form:
\begin{align}
\langle e^{\cal Q}\, e^{-\be H}\rangle_N
\,\simeq\,{\CN}^{-2} (e^{i{\bth}_N/2})  \sum\limits_{{\blad^{L, R}}
\subseteq \{\CK^N\}}
S_{{\blad^L}}(e^{-i{\bth}_N})\,
S_{{\blad^R}}(e^{i{\bth}_N})
& \nonumber
\\ \times\,
\exp\Bigl(\sum_{k=1}^{N}\al_{\mu^L_k}
\Bigr)\,
G_{{\bmu^L}; {\bmu^R}} (\be) \,,&
\label{cor:lin668}
\end{align}
where
\begin{equation}
G_{{\bmu^L}; {\bmu^R}} (\be) \, \simeq \,e^{\be h (\frac{M}2 - N)} \times \det\bigl(I_{|{ {\mu_{i}^L}- {\mu_{j}^R}}|}(\be)\bigr)_{1
\le i, j \le N}\,.
\label{cor:lin668888}
\end{equation}

\begin{figure}[h]
\center
\includegraphics [scale=0.9] {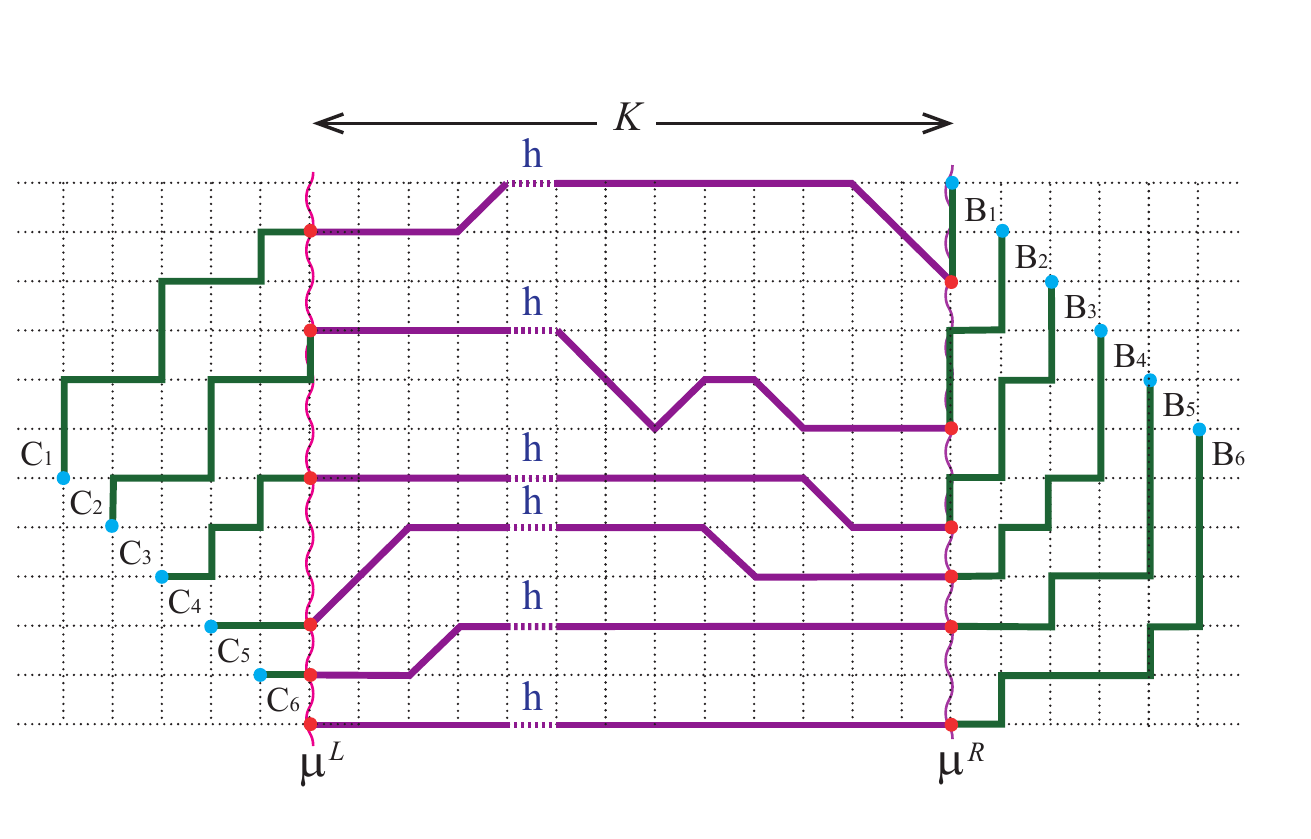}
\caption{Nest of paths contributing to $\mathfrak{P}(\textbf{1}_N; \textbf{1}_N | K)$ at $N=6$, $K=13$, and $p=1$. }
\label{fig:f9}
\end{figure}

As it follows from \textsf{Proposition~6}, the representation (\ref{cor:lin668}),  (\ref{cor:lin668888}) is related to superposed random walks (cf.~\cite{nest}). Indeed,
applying $\lim\limits_{\{\al_k \to 0\}}\,\cd^{\,l}_{\al_1, \al_2,\dots ,
\al_{l}}$
to the nominator of (\ref{cor:lin665})
taken over the ground state solution (\ref{grstxx}), one obtains, with the use of
(\ref{cor:lin668}), the generating function of self-avoiding lattice paths of special type:
\begin{equation}
\mathcal D^K_{\be/2}\,
\langle \Psi (e^{i{\bth}^{\rm g}_N/2}) \Bigl| \prod_{i=1}^{l} {\sf q}_{i}\, e^{- \be H} \Bigr|
\Psi (e^{i{\bth}^{\rm g}_N/2})
\rangle\,=\, \mathfrak{P}(e^{i{\bth}^{\rm g}_N/2}; e^{i{\bth}^{\rm g}_N/2}\,| K)\,.
\label{cor:linn5571}
\end{equation}
The number $\mathfrak{P}(e^{i{\bth}^{\rm g}_N/2}; e^{i{\bth}^{\rm g}_N/2}\,| K)$ in right-hand side of (\ref{cor:linn5571}) is due to the substitute ${\textbf u}_N = {\textbf v}_N = e^{i{\bth}^{\rm g}_N/2}$ in the polynomial
\begin{equation}
\label{cor:linnn5571}
\mathfrak{P}({\textbf v}_N; {\textbf u}_N\,| K) \equiv
\sum_{\w\blad^L, \blad^R \subseteq \{{\CK}^N\}} S_{\w\blad^L} ({\textbf v}_N^{-2}) S_{\blad^R} ({\textbf u}_N^{2})\,
\mathfrak{G}({\w\bmu^L}; {\bmu^R}\,| K)\,,
\end{equation}
where $\sum_{\w\blad^L \subseteq \{{\CK}^N\}}$
goes over $\w\blad^L$ (\ref{cor:lin5581}), and
$\mathfrak{G}({\w\bmu^L}; {\bmu^R}\,| K)$ is given by (\ref{mpcf4}). The replacement
$e^{i{\bth}^{\rm g}_N}\longmapsto 1$ is appropriate at $M\gg N$, and one obtains from (\ref{cor:linn5571}):
\begin{equation}
\mathcal D^K_{\be/2}\,
\langle \Psi(\textbf{1}_N) \Bigl| \prod_{i=1}^{l} {\sf q}_{i}\, e^{- \be H} \Bigr| \Psi(\textbf{1}_N)
\rangle\,=\, \mathfrak{P}(\textbf{1}_N; \textbf{1}_N | K)\,,
\label{cor:linnn05571}
\end{equation}
where
\begin{align}\label{cor:linnn15571}
\mathfrak{P}(\textbf{1}_N; \textbf{1}_N
| K) &=\,\sum_{i=0}^{K}
\begin{pmatrix}
K\\
i
\end{pmatrix}
\bigl(h(M-2N) \bigr)^i\, \sum_{|{\bf n}|=K-i} P({\bf n})\,
{\bar{\bold\Delta}}^{\bf n}\,,\\
\label{cor:linnn15572}
{\bar{\bold\Delta}}^{\bf n} &\equiv\,
\sum_{\w\blad^L, \blad^R \subseteq \{{\CK}^N\}} S_{\w\blad^L} (\textbf{1}_N)
S_{\blad^R} (\textbf{1}_N)
\det\bigl(({\bold\Delta}^{n_j} )_{\w\mu^L_{i}; \mu^R_j}\bigr)_{1
\le i, j \le N}\,,
\end{align}
and $({\bold\Delta}^{n_j} )_{\w\mu^L_{i}; \mu^R_j}$ is given by (\ref{qanal276}).
The Schur polynomials are related to the nests of lattice paths, Fig.~\ref{fig:f5}, and therefore the polynomials (\ref{cor:linnn5571}) are related  to the nests of lattice paths of the type in Fig.~\ref{fig:f9}. It is seen from (\ref{mpcf5}) that
$\mathfrak{P}(\textbf{1}_N; \textbf{1}_N
| K)$ (\ref{cor:linnn15571}) are
the polynomials of two variables, $h(M-N)$ and $-hN$, with integer coefficients
related to enumeration of self-avoiding lattice paths. A typical term of the sum (\ref{cor:linnn15572}) is depicted in Fig.~\ref{fig:f9} for $K=13$ and $p=1$,
so that ${\bar{\bold\Delta}}^{\bf n}$ (\ref{cor:linnn15572}) is the number of the nests of paths characterized by ${\bf n}=(0, 1, 3, 1, 4, 3)$, $|{\bf n}|= 12$, while all admissible ``crossings'' with the dissection lines occur.

\section{The generating function $G({\bf a}_M)$ and the correlation functions of flipped spins}
\label{sec60}
\subsection{The $N$-particles mean values at large length of the chain}

Let ${\sf trace}$ in (\ref{cor:lin5}) be the trace over all $N$-particles Bethe states, and let us consider the $N$-particles trace of the Boltzmann-weighted generating exponential:
\begin{equation}
\label{ratbee7003}
\tr_N(e^{\cQ} e^{-\be H})\,\equiv\,
\sum_{\{{\bth}_N \}} \langle e^{\cQ}\, e^{-\be H} \rangle_N\,,
\end{equation}
where $\sum_{\{{\bth}_N\}}$ denotes summation over independent $N$-particles solutions to (\ref{betheexp}).
The definition (\ref{ratbee7003})
enables to define the $N$-particles mean value: \begin{equation}
{{\langle\langle}} e^{\cQ} {{\rangle\rangle}}_{\be, N}\,
\equiv\, \tr_N(e^{\cQ}\boldsymbol{\rho}_N )\,,\qquad \boldsymbol{\rho}_N \equiv \frac{e^{-\be H}}{\tr_N(e^{-\be H})}
\,.
\label{new11}
\end{equation}
We express (\ref{new11}) using (\ref{cor:lin665}):
\begin{equation}\label{new111}
{{\langle\langle}} e^{\cQ} {{\rangle\rangle}}_{\be, N}\,
=\, \frac{\sum\limits_{\{{\bth}_N \}}
\det ( e^{- \be {\h\ep}} {e^{\h\al}}) }{\sum\limits_{\{{\bth}_N \}} \det ( e^{- \be {\h\ep}})}\,.
\end{equation}

In order to investigate (\ref{new111}) at large $M\gg 1$,
it is more appropriate to evaluate (\ref{ratbee7003}) using
the integral representation (\ref{cor:lin668}), (\ref{cor:lin668888}):
\begin{equation}
\tr_N(e^{\cQ} e^{-\be H}) = \sum\limits_{\{{\bmu}_N\}}
\exp\Bigl(\sum_{k=1}^{N}\al_{\mu_k}
\Bigr)\,G_{{\bmu}; {\bmu}} (\be) \label{ratbe7003}\,,
\end{equation}
where (\ref{ratbe77}) is taken into account to sum up over the sets of the Bethe solutions, and $\sum_{\{{\bmu}_N\}}$ is to re-express, for convenience, $\sum_{{\blad}
\subseteq \{\CK^N\}}$. The mean value (\ref{new11}) is estimated with the use of (\ref{ratbe7003}):
\begin{equation}
{{\langle\langle}} e^{\cQ} {{\rangle\rangle}}_{\be, N}\Big|_{M\gg 1}\,
\simeq\,\frac{{\cal I}_N (\be, {\bf a}_{M})}{{\cal I}_N (\be, {\bf 0}_{M})} \,,
\label{new12}
\end{equation}
where
\begin{equation}
\label{ratbe711432}
{\cal I}_N (\be, {\bf a}_{M}) \equiv
\sum_{\{{\bmu}_N\}} \det \bigl(e^{\al_{\mu_i}} I_{|\mu_i-\mu_j|}(\be) \bigr)_{1\le i, j\le N}\,.
\end{equation}
With regard at {\sf Proposition~6}, the following power series is valid for ${\cal I}_N (\be, {\bf a}_{M})$ (\ref{ratbe711432}):
\begin{equation}
\label{ratbe71131}
{\cal I}_N (\be, {\bf a}_{M})\,
=\,\sum_{K=0}^{\infty}
\frac{(\be/2)^K}{K!}\,\sum_{|{\bf n}|=K} P({\bf n})\,{\bold\Delta}^{\bf n}({\bf a}_{M})\,,
\end{equation}
where
\begin{equation}
{\bold\Delta}^{\bf n}({\bf a}_{M}) \equiv\,
\sum_{\{{\bmu}_N\}} \det\left(
e^{\al_{\mu_j} } \begin{pmatrix}{n_j} \\ \frac{{n_j}+\mu_j-\mu_i}2
\end{pmatrix}\right)_{1
\le i, j \le N}\,.
\end{equation}

Let us consider the parametrization ${\bf a}_M = \al {\bf M}$, where ${\bf M} = (1, 2, \ldots, M)$ (see (\ref{new30})). Then, we re-express
(\ref{ratbe711432}):
\begin{equation}
\label{new15}
{\cal I}_N (\be, \al {\bf M})\,=\,
\sum_{\{{\bmu}_N\}}
e^{\al |{\bmu}|} \det \bigl( I_{|\mu_i-\mu_j|}(\be) \bigr)_{1\le i, j\le N}\,.
\end{equation}
Applying $\mathcal D^l_{\al}$ to
(\ref{new12}) and (\ref{new15}), we obtain:
\begin{equation}
{{\langle\langle}} {\sf M}^l {{\rangle\rangle}}_{\be, N}\Big|_{M\gg 1}\,
\simeq\,\frac{\sum_{\{{\bmu}_N\}}
|{\bmu}_N|^l \det \bigl( I_{|\mu_i-\mu_j|}(\be) \bigr)_{1\le i, j\le N}}{{\cal I}_N (\be, {\bf 0}_{M})} \,,
\label{new17}
\end{equation}
where ${\sf M}$ is defined in (\ref{new30}). Moreover, Eq.~(\ref{new15}) is telling that
\begin{equation}
\mathcal D^l_{\al}\,\mathcal D^K_{\be/2}\,{\cal I}_N (\be, \al {\bf M})\,
=\,\sum_{\{{\bmu}_N\}} |{\bmu}_N|^l |P^0_{K} ({\bmu}_N \rightarrow\,{\bmu}_N)|\,.
\label{new16}
\end{equation}
Right-hand side of (\ref{new16})
may be viewed as the sum of the numbers
\begin{equation}
\label{new18}
n^l\,\sum_{{\bmu}_N \vdash n} |P^0_{K} ({\bmu}_N \rightarrow\,{\bmu}_N)|\,,
\end{equation}
where such sets of ``closed'' trajectories are summed up that the initial ($\equiv$ final) pos\-it\-ions of vicious walkers constitute a partition of appropriate $n\in\BN$.

For a given $l$-tuple ${\bf k}_l$, let $\{{\w\bmu_N}\}_{{\bf k}_l}$ be
the set of all admissible strict partitions of length $N$, which contain $l$ parts of ${\bar{\bf k}_l}$. With regard at (\ref{cor:lin611}) and (\ref{cor:lin61}), we obtain from (\ref{new12}):
\begin{equation}
{{\langle\langle}} {\varPi}_{\bf k}\, {{\rangle\rangle}}_{\be, N}\Big|_{M\gg 1}\,
\simeq\,\frac{\w{\cal I}_N (\be, {\bf k}_l)}{{\cal I}_N (\be, {\bf 0}_{M})} \,,
\label{new13}
\end{equation}
where
\begin{equation}
\w{\cal I}_N (\be, {\bf k}_l)\,
\equiv\,
\sum_{\{{\w\bmu}_N\}_{{\bf k}_l}} \det \bigl(I_{| \w\mu_i - \w\mu_j|}(\be) \bigr)_{1\le i, j\le N}\,,
\label{ratbe711433}
\end{equation}
and
$\sum_{\{{\w\bmu_N}\}_{{\bf k}_l}} \equiv \sum_{\{{\w\bmu_N}\}}$ is the sum as in (\ref{55441}). Furthermore, {\sf Proposition~6} tell us that
\begin{align}
\label{ratbe71143}
\w{\cal I}_N (\be, {\bf k}_l) &=\,\sum_{K=0}^{\infty}
\frac{(\be/2)^K}{K!}\,\sum_{|{\bf n}|=K}
P({\bf n}) {\w
{\bold\Delta}}^{\bf n}({\bf k}_l)\,,\\
\label{ratbe711431}
{\w{\bold\Delta}}^{\bf n}({\bf k}_l) &\equiv \sum_{\{{\w\bmu}_N\}_{{\bf k}_l}} \det\left( \begin{pmatrix}{n_j} \\ \frac{{n_j}+{\w\mu}_j-{\w\mu}_i}2
\end{pmatrix} \right)_{1
\le i, j \le N}\,.
\end{align}

Let us introduce
the number of all sets of trajectories of $N$ random turns vicious walkers initially located at ${\w\bmu_N}\in \{{\w\bmu_N}\}_{{\bf k}_l}$ and returning after $K$ steps to their initial positions:
\begin{equation}
|{\cal P}^0_{K, N} ({\bf k}_l)| \equiv \sum_{\{{\w\bmu}_N\}_{{\bf k}_l}} |P^0_{K} ({\w\bmu}_N \rightarrow\,{\w\bmu}_N)|\,,
\label{ratbe711438}
\end{equation}
where $|P^0_{K} ({\w\bmu}_N \rightarrow\,{\w\bmu}_N)|$ is
defined by (\ref{qanal272}). Due to (\ref{ratbe71143}), the
function $\w{\cal I}_N (\be, {\bf k}_l)$ (\ref{ratbe711433}) is the generating function of the numbers $|{\cal P}^0_{K, N} ({\bf k}_l)|$ (\ref{ratbe711438}):
\begin{equation}
\mathcal D^K_{\be/2}\, \w{\cal I}_N (\be, {\bf k}_l)\,
=\,|{\cal P}^0_{K, N} ({\bf k}_l)|\,.
\label{new14}
\end{equation}

\subsection{Determinantal representation of $G ({\bf a}_M)$}
\label{sec601}

Let us proceed with the evaluation of $G({\bf a}_M)$
(\ref{cor:lin5}), where ${\sf trace}$ is defined conventionally, \cite{KBI2, col1, col2}, and includes summation over sets of Bethe solutions and over numbers of particles:
\begin{align}
\Tr\bigl(e^{\cQ}\, e^{-\be H}\bigr) & =\,\sum_{N=0}^{M}
\sum_{\{{\bth}_N\}} \langle e^{\cQ}\, e^{-\be H} \rangle_N
\nonumber
\\
\label{cor:lin669}
& = e^{\be h M/2}\,\Bigl(1+\sum_{N=1}^{M}
\sum_{\{{\bth}_N\}} \det
(e^{- \be {\h\ep}} {e^{\h\al}})\Bigr)\,.
\end{align}
Equation (\ref{cor:lin665}) is used in (\ref{cor:lin669}), and the averaging at $N=0$ is over $\mid\Uparrow\rangle$. Besides,  $Z=\Tr\bigl(e^{-\be H}\bigr)$ (\ref{cor:lin5}) results from (\ref{cor:lin669}) at $\h\al = 0$.

Taking into account the definition (\ref{cor:linn23}),
one transforms (\ref{cor:lin669}):
\begin{align}
\label{tr2}
\Tr\bigl(e^{\cQ}\, e^{-\be H} \bigr)\,=\,\frac{e^{\be h M/2}}{2}\sum_{\ell=\pm 1}
\bigl({\sf D}_{+}^{(\ell)}({\h\al})+\ell\,
{\sf D}_{-}^{(\ell)}({\h\al}) \bigr)\,,
\\
\label{tr1}
{\sf D}_{\pm}^{(\ell)}({\h\al}) \equiv \det({\h I}+ \ell e^{- \be {\h\ep}} {e^{\h\al}})_{p\in {\sf S}^\pm}
\,, \qquad \ell=\pm 1\,,
\end{align}
where the subscript $p\in {\sf S}^\pm$ reminds that the entries of $M\times M$ matrices $e^{\h\al}$ and $e^{- \be {\h\ep}}$ are parameterized by elements of  ${\sf S}^\pm$ (\ref{cor:linn23}) (cf.~(\ref{cor:lin55311}) and (\ref{mpcf1}); for instance, $\h\ep\equiv {\diag} \{\ep_p\}_{p\in {\sf S}^\pm}$, where $\ep_p\equiv \ep (p)$ is given by (\ref{egen})), and $\h I$ is unit $M\times M$ matrix. The identity (\ref{tr2}) is verified provided that the Laplace formula for determinant of sum of two matrices is applied \cite{KBI2}.

Further, let us consider the following determinantal identities:
\begin{equation}
\label{cor:lin674}
{\sf D}_{\pm}^{(\ell)} ({\h\al}) =
{\sf G}_{\pm}^{(\ell)}\,
{\sf D}_{\pm}^{(\ell)}({\h\al=0})\,,\quad
{\sf G}_{\pm}^{(\ell)} \equiv \det({\h I}\,+ {\h M}_{\rm xx}^{(\ell)})_{p\in {\sf S}^\pm}\,,
\end{equation}
where the matrix ${\h M}_{\rm xx}^{(\ell)}$ is defined:
\begin{equation}
{\h M}_{\rm xx}^{(\ell)}\,\equiv\, ({{e^{\h\al}}}-{\h I}){\h f}^{(\ell)}\,,\qquad
{\h f}^{(\ell)} \equiv
({\h I} + \ell\,e^{\be
{\h\ep}})^{\1}\,.
\label{cor:lin67002}
\end{equation}
The determinantal representation for $G ({\bf a}_M)$ (\ref{cor:lin5}) resulting from Eqs.~(\ref{tr2}) and
(\ref{cor:lin674}) is reduced, under the conventional specification of ${\bf a}_M\ni \al_n$ (cf.~(\ref{cor:lin55311})),
to the average $\l\l e^{\al Q(m)} \r\r_\be$ derived in \cite{col2}. It is seen from (\ref{tr2}) that the limiting form of $G ({\bf a}_M)$ at growing $M$
is due to $\ell=+1$
whereas the terms at $\ell=-1$ are mutually cancelled as soon as ${p\in {\sf S}^\pm}$ is replaced by ${p\in {\sf S}}$. Therefore, $\l\l e^{\al Q(m)} \r\r_\be$ becomes the Fredholm determinant at $M\to\infty$: the matrices are replaced by appropriate kernels, the integration arises instead of the matrix multiplication, etc., \cite{col}. The same
is expected for the determinantal
representation of $G({\bf a}_M)$. However, additional requirement $\lim_{n\to\infty} \al_n = 0$ has to be imposed here. Since the interest to $G({\bf a}_M)$ is rather motivated by its role of the generating function, we shall not pay attention to $G({\bf a}_M)$ as the Fredholm determinant.

Using (\ref{ratbe7003}) and (\ref{cor:lin669}), one arrives at the following

\vskip0.3cm \noindent
\noindent{\bf Statement~2:\,}

\noindent\textit{$\bullet\,\,$
Total trace of the Boltzmann-weighted generating exponential is represented at large enough $M$}:
\begin{equation}
\label{ratbe7103}
\Tr\bigl(e^{\cQ}\, e^{-\be H} \bigr)\,\simeq\, e^{\be h M/2} \Bigl(1+ \sum_{N=1}^{M\gg 1} e^{-\be h N} {\cal I}_N (\be, {\bf a}_{M})\Bigr)\,,
\end{equation}
\textit{where ${\cal I}_N (\be, {\bf a}_{M})$ is given by} (\ref{ratbe711432}).

\vskip0.3cm
\noindent\textit{$\bullet\,\,$ The mean value of ${\varPi}_{\bf k}$ defined by \eqref{cor:lin611}, \eqref{cor:lin61} acquires, with regard at \eqref{ratbe711432}, \eqref{ratbe7103} the ratio form}
\begin{equation}
\l\l {\varPi}_{\bf k} \r\r_\be = \frac{\varPhi (\be, h, {\bf k}_l)}{Z}\,,
\label{ratbe711437}
\end{equation}
\textit{where}
\begin{align}
\varPhi (\be, h, {\bf k}_l)  & \equiv \lim\limits_{\{\al_k \to 0\}}\,
\cd^{\,l}_{k_1, k_2,
\dots , k_{l}} \Tr\bigl(e^{\cQ}\, e^{-\be H} \bigr) \nonumber \\
& \simeq
e^{\be hM/2} \sum_{N=l}^{M\gg 1}
e^{-\be h N} \w{\cal I}_N (\be, {\bf k}_l)\,.
\label{ratbe7114}
\end{align}

\vskip0.3cm \noindent
The partition function $Z=\Tr\bigl(e^{-\be H}\bigr)$ (\ref{cor:lin5}) arises from  (\ref{ratbe7103}) provided that
${\bf a}_M$ consists of zeros, and $Z$ is expressed, due to {\sf Proposition~6}, through the numbers $|P^0_K ({\bmu}_N \rightarrow\,{\bmu}_N)|$.

With regard at
(\ref{mpcf4}), we define the polynomial the coefficients of which are the numbers of sets of trajectories with staying of $N$ walkers admitted
(typical set is shown in Fig.~\ref{fig:f1}):
\begin{equation}
{\cal P}_{K, N} ({\bf k}_l) \equiv
\sum_{i=0}^{K}
\begin{pmatrix}
K\\
i
\end{pmatrix}
\bigl(h(M-2N) \bigr)^i\,
|{\cal P}^0_{K-i, N} ({\bf k}_l)|\,,
\nonumber
\end{equation}
where $|{\cal P}^0_{K-i, N} ({\bf k}_l)|$ are defined by (\ref{ratbe711438}). Therefore, $\varPhi (\be, h, {\bf k}_l)$ (\ref{ratbe7114}) plays the role of the generating function of the polynomials encoding the total number of all sets of ``closed'' trajectories of random turns walkers such that $l$ their initial/final positions coincide (for each $N$) with the sites ${\bf k}_l$:
\begin{equation}
\nonumber
\mathcal D^K_{\be/2} \varPhi (\be, h, {\bf k}_l) \equiv \sum_{N=l}^{M}
{\cal P}_{K, N} ({\bf k}_l)\,.
\end{equation}

\subsection{Differentiation of $G ({\bf a}_M)$}

Let us consider differentiation of the generating function $G ({\bf a}_M)$. We introduce the shortening notations ${\sf G} \equiv {\sf G}_{\pm}^{(\ell)}$, ${\h R}\equiv (\h I +{\h M}_{\rm xx}^{(\ell)})^{\1}$, and obtain  the first order derivative:
\begin{equation}
{\sf G}^{\1}\cd_{k_1} {\sf G}\,=\, e^{\al_{k_1}} \tr\big({\h {R}}\, \h\dl_{k_1} {\h f} \big)\,=\,
e^{\al_{k_1}}
{\bar R}_{k_1, k_1} \,,
\label{cor:bin2}
\end{equation}
where $\cd_l\equiv\cd/\cd\al_{l}$ and ${\h\dl}_{l}\equiv \cd_{l} {\h\al}$. The diagonal matrix ${\h f}\equiv {\h f}^{(\ell)}$ is given by (\ref{cor:lin67002}), and ${\bar R}_{k_1, k_1}$ is the diagonal entry of the matrix
${\bar R}\equiv \{{\bar R}_{m n}\}_{1\le m, n\le M}$, where
\begin{equation}
{\bar R}_{m n}\,\equiv\, \frac1M \sum_{p, q} e^{-i n p} f_p\,{R}_{p q} e^{i m q} \,,
\label{cor:bin4}
\end{equation}
and summation is over sets (\ref{cor:linn23}) appropriately.
The second order derivative
of ${\sf G}$ is obtained,
\begin{eqnarray}
&& {\sf G}^{\1} \cd^2_{k_1, k_2} {\sf G}\,=\,e^{\al_{k_1}+ \al_{k_2}}
\left|\,\begin{matrix} {\bar R}_{k_1, k_1} & {\bar R}_{k_1, k_2}
\\ {\bar R}_{k_2, k_1} & {\bar R}_{k_2, k_2} \end{matrix}\, \right|\,,
\label{cor:bin5}
\end{eqnarray}
since $\tr\big({\h
R}\,\h \dl_{k_1} {\h f} {\h
R}\,\h\dl_{k_2} {\h f} \big)$
takes the product form ${\bar R}_{k_1, k_2} {\bar R}_{k_2, k_1}$ due to (\ref{cor:bin4}).

With regard at (\ref{cor:bin2}) and (\ref{cor:bin5}), one formulates the following

\noindent{\bf Proposition~7:} \textit{The function ${\sf G}$ defined by \eqref{cor:lin674} is the generating function of the minors of the matrix} ${\bar R}$  (\ref{cor:bin4}),
\begin{equation}
{\sf G}^{\1} \cd^{\,l}_{k_1, k_2,\dots ,
k_{l}} {\sf G}\,=\, e^{\al_{k_1} + \al_{k_2} + \ldots + \al_{k_l}}\,{\det}_l {\sf R}\,,
\label{cor:bin6}
\end{equation}
\textit{where $1 \le k_1 < k_2 < \ldots < k_l \le M$, $\cd^{\,l}_{k_1, k_2,\dots ,
k_{l}}$ is defined by \eqref{cor:lin61}, and ${\det}_l
{\sf R}$ is the minor given by the submatrix of $l^{\rm th}$ order
$\{{\sf R}_{i j}\}_{1 \le i, j  \le l} \equiv \{{\bar R}_{k_i, k_j}\}_{1 \le i, j \le l}$}.

\noindent{\bf Proof:} We use induction with the base case (\ref{cor:bin2}) and induction step consisting in validity of (\ref{cor:bin6}) at $l-1$,
\begin{equation}
{\sf G}^{\1} \cd^{\,l-1}_{k_1, k_2,\dots ,
k_{l-1}} {\sf G}\,=\,e^{\al_{k_1} + \al_{k_2}+\ldots + \al_{k_{l-1}}}\, {\det}_{l-1} {\sf R} \,.
\label{cor:bin8}
\end{equation}
Then, the relation
$\cd_{\al_{k_n}} {{\bar R}}_{k_i, k_j}\,=\,e^{\al_{k_n}} {{\bar R}}_{k_i, k_n}
{{\bar R}}_{k_n, k_j}$ leads from (\ref{cor:bin8}) to
\begin{equation}
\cd^{\,l}_{k_1, k_2,\dots ,
k_{l-1}, k_{l}} {\sf G}\, = \,e^{\al_{k_1} + \al_{k_2}+ \ldots + \al_{k_{l-1}}}\, \bigl({\sf G}\, \cd_{k_{l}} {\det}_{l-1} {\sf R}\,+\,{\det}_{l-1} {\sf R}\, \cd_{k_{l}} {\sf G}\bigr)\,.
\label{cor:bin9}
\end{equation}
The derivative of ${\det}_{l-1} {\sf R}$ is of the form:
\begin{equation}
\cd_{k_{l}}{\det}_{l-1} {\sf R}\,=\,\sum \limits^{l-1}_{i=1} (-1)^{l+i} {\sf R}_{l i}\,
\left|\,\begin{matrix} {\sf R}_{1 1} &
{\sf R}_{1 2} & \dots &
\check {\sf R}_{1 i} & \dots & {\sf R}_{1 l}
\\ {\sf R}_{2 1} & {\sf R}_{2 2} & \dots &
\check {\sf R}_{2 i} & \dots & {\sf R}_{2 l}\\
\dots & \dots & \dots &\dots &\dots &\dots \\
{\sf R}_{l-1, 1} & {\sf R}_{l-1, 2} & \dots &
\check {\sf R}_{l-1, i} & \dots & {\sf R}_{l-1, l}
 \end{matrix}\, \right|\,,
\label{cor:bin10}
\end{equation}
where $\check {\sf R}_{i j}$ implies that the relevant column is omitted. The main statement (\ref{cor:bin6})
arises from (\ref{cor:bin9})
due to (\ref{cor:bin2}) and (\ref{cor:bin10}). $\Box$

\noindent $\bullet\,$ {\sf Proposition~7} is telling us that the average (\ref{cor:lin611}) on infinite chain
takes the determinantal form since ${\sf G}$ tends to unity at $\al_n \rightarrow 0$, $\forall n$:
\begin{equation}
\l\l {\varPi}_{\bf k} \r\r_{\be} = \lim_{M\to\infty}
\lim \limits_{\{\al_k \to 0\}}\,\cd^{\,l}_{k_1, k_2,\dots ,
k_{l}} {\sf G}_{\pm}^{(\ell)}\, =\,{\det} \bigl(
{f}_{k_i, k_j} \bigr)_{1 \le i, j \le l}\,,
\label{cor:bin11}
\end{equation}
where the entries ${f}_{k_i, k_j}$ are given by
(\ref{cor:bin4}) with respect of the fact that $\h R$ tends to unit matrix. Therefore, the limiting relation is valid for
${Z}^{-1} \varPhi (\be, h, {\bf k}_l)$ (\ref{ratbe711437}):
\begin{equation}\label{new92}
\lim_{M\to\infty} \frac{\varPhi (\be, h, {\bf k}_l)}{Z}\,=\,{\det} \bigl(
{f}_{k_i, k_j} \bigr)_{1 \le i, j \le l} \,.
\end{equation}

\subsection{The asymptotics at increasing $\be$}

The representation $\tr_N(e^{\cQ} e^{-\be H})$ (\ref{ratbe7003}) can be estimated
at $1\ll M \ll \be$ as follows.
Now Eq.~(\ref{cor:lin667}) is used to sum up over the sets of the Bethe solutions, and one obtains:
\begin{align}
\nonumber
\frac{\tr_N(e^{\cQ} e^{-\be H})}{e^{\be h M/2}}\,=\,
\frac{1}{N!} \int_{I_N} {\cal P}_{\CK}
(e^{-i{\bf p}_N}, e^{i{\bf p}_N}, {\bf a}_M)\,| \CV (e^{i{\bf p}_N}) |^2 \\
\label{ratbe7113}
\times\,e^{{\be}\sum_{l=1}^N
(\cos{p}_l - h)}
\frac{d^Np}{(2\pi)^N}
\,,
\end{align}
where ${\cal P}_{\CK}$ is the sum (\ref{cor:lin552}).
We approximate (\ref{ratbe7113})
at $\be\gg 1$:
\begin{align}
\frac{\tr_N(e^{\cQ} e^{-\be H})}{e^{\be h M/2}}\,&\simeq\, {\cal P}_{\CK} ({\bf 1}_N, {\bf 1}_N, {\bf a}_{M})\,
V_N(\be, h)\,,
\label{ratbe072} \\
V_N(\be, h)\,&\equiv\,
\frac{e^{\be N(1-h)}\, \mathfrak{I}_N}{\be^{N^2/2}}\,=\,
e^{\be N(1-h)-\frac{N^2}{2}\log\be + \varphi_N}\,,\quad \varphi_N\equiv \log \mathfrak{I}_N \,,
\label{ratbe791}
\end{align}
where ${\cal P}_{\CK} ({\bf 1}_N, {\bf 1}_N, {\bf a}_{M})$ arises at $q\to 1$ from (\ref{cor:lin552}) under the $q$-parametrization (\ref{rep21}). Furthermore,
$\mathfrak{I}_N$ in (\ref{ratbe791}) is Mehta integral \cite{meh},
\begin{equation}
\nonumber
\mathfrak{I}_N\,\equiv\,
\displaystyle{\frac{1}{N!}
\prod_{k=1}^{N}
\Bigl(\int\limits_{-\infty}^{\infty}
\frac{d p_k}{2\pi} \Bigr)
e^{\frac{-1}2 \sum\limits_{i=1}^{N} {p}^2_i}} \prod_{1\leq k<l \leq N} \bigl|{p}_k - {p}_l \bigr|^2\,,
\end{equation}
which is expressed in terms of the Barnes $G$-function \cite{barn}:
\begin{equation}
\nonumber
\mathfrak{I}_N\,=\, \frac{G(N+1)}{(2\pi)^{N/2}}\,,
\qquad G(N+1)\,\equiv\,\frac{(N !)^N}{1^1\,2^2\,\ldots N^N }\,=\,
\prod^N_{k=1} \Gamma(k)\,.
\end{equation}
The behaviour of $\mathfrak{I}_N = e^{ \varphi_N}$ at $M\gg N\gg 1$ is due to the following estimate of $\varphi_N$ \cite{bmumn}:
\begin{equation}
{{\varphi}}_N\,=\,\frac{N^2}{2} \log N\,-\,
\frac{3 N^2}{4}\,+\,{\cal O}(\log
N)\,,\qquad N\gg 1\,. \label{spdfpxx6}
\end{equation}
From (\ref{spdfpxx6}) it is seen that $V_N(\be, h)$ (\ref{ratbe791}) depends on $\frac{\be}{N}$ at $\be> N\gg 1$ appropriately for an opportunity of the third order phase transition \cite{tier, zah}.

Provided that the specification $\al_n= n \log \ga$, $0<\ga\le 1$ (cf. Section~\ref{sec53}) is adopted, the values ${\cal P}_{\CK}({\bf 1}_N, {\bf 1}_N, {\bf a}_{M})$ in (\ref{ratbe072}) arise due to the limit $q\to 1$ in
\begin{align}
\nonumber
{{\cal P}}_{\CK}
\Bigl(\textbf{q}_N, \frac{\textbf{q}_N}{q}, {\bf a}^\ga_{M}\Bigr) & =\,\langle e^{\cQ (\ga)}
\rangle_{N,q}\,= \\
\label{ratbe721}
& =\,\ga^{\frac{N}{2} (N+1)}\,{G}(N, N, {\CM} |\,q, \ga)
\,,
\end{align}
where ${\bf a}^\ga_{M} \equiv \log \ga\cdot (1, 2, \ldots, M)$, and $\langle e^{\cQ (\ga)}
\rangle_{N,q}$
defined by (\ref{ratbee707})
is given by (\ref{cor:lin559}), (\ref{cor:lin5591}). The behaviour at large enough $M$ is approximately given by the limiting expression (\ref{cor:lin5597}) at $q\to 1$:
\begin{align}
\nonumber
{\cal P}_{\CK}({\bf 1}_N, {\bf 1}_N, {\bf a}^\ga_{M}) )\Big|_{M\gg 1} & =\,\ga^{\frac{N}{2} (N+1)}\,{G}(N, N, {\CM} |\,1, \ga)\Big|_{M\gg 1}
\\
\nonumber
& \underset{M\to\infty} \longrightarrow\, \ga^{\frac{N}{2} (N+1)}
\lim_{q\to 1}\,
\displaystyle{
\prod_{i=1}^{N} \prod_{j=1}^{N} \frac{1}{1- \ga
q^{i+j-1} }}\,.
\end{align}
Here, ${G}(N, N, {\CM} |\,1, \ga)\Big|_{M\gg 1}$ is the generating function of the number of plane partitions with fixed sum of its diagonal elements confined in $N\times N\times {\CM}$ box at $M\gg 1$.

Using (\ref{ratbe721}), we obtain the limiting value of the $N$-particles mean value of the generating exponential (\ref{new11}):
\begin{equation}
\label{ratbe723}
{{\langle\langle}} e^{\cQ(\ga)} {{\rangle\rangle}}_{\be, N} \Bigl|_{1\ll M\ll\be}\,\simeq\,
\frac{{G}(N, N, {\CM} |\,1, \ga)\Big|_{M\gg 1}}{{A}(N, N, {\CM})\Big|_{M\gg 1} }\,.
\end{equation}
According to (\ref{ratbe794}) and (\ref{cor:llin5591}), the denominator in (\ref{ratbe723}) is the number of plane partitions confined in $N\times N\times {\cal M}$ box with increasing height. Let us remind that, according to (\ref{cor:lin559}), the generating function ${G}(N, N, {\CM} |\,1, \ga)$ (\ref{ratbe707}) is of a polynomial form. Therefore,
the differentiation gives:
\begin{align}
\nonumber
{{\langle\langle}} {\sf M}^l {{\rangle\rangle}}_{\be, N}\Big|_{1\ll M\ll\be} &
\simeq\,\frac{\mathcal D^l_{\al} {G}(N, N, {\CM} |\,1, e^{\al})\Big|_{M\gg 1}} {{A}(N, N, {\CM})\Big|_{M\gg 1}} \\
\label{ratbbbe723}
& =\,\frac{\sum_{\{{\bmu}_N\}} |{\bmu}_N|^l
S_{\blad} (\mathbf{1}_N) S_{\blad} (\mathbf{1}_N)
\Big|_{M\gg 1}}{{A}(N, N, {\CM})\Big|_{M\gg 1} }\,. \end{align}
The nominator in right-hand side of (\ref{ratbbbe723}) may be viewed as a sum of the numbers
\begin{equation}
\label{new40}
n^l {A}(N, N, {\CM} |\,n)\,,
\quad {A}(N, N, {\CM} |\,n) \equiv \sum_{{\bmu}_N \vdash n}
S_{\blad} (\mathbf{1}_N) S_{\blad} (\mathbf{1}_N)
\end{equation}
where ${A}(N, N, {\CM} |\,n)$
denotes the number of plane partitions with $\tr_N{\bpi} = n$ confined in the corresponding $N\times N\times {\cal M}$ box. Right-hand side of (\ref{ratbbbe723}) is less than unity, and it temptingly tends to zero at $M\to\infty$.

In the case of the mean value of the projector ${\varPi}_{\bf k}$, one obtains:
\begin{equation}
\label{ratbe724}
{{\langle\langle}} {\varPi}_{\bf k} {{\rangle\rangle}}_{\be, N}
\Bigl|_{1\ll M\ll\be}\,\simeq\,
\frac{{\w {\cal P}}_{\CK}
(\textbf{1}_N, \textbf{1}_N, {\bf k}_l)}{{ {\cal P}}_{\CK}
(\textbf{1}_N, \textbf{1}_N, {\bf 0})}\,=\,\frac{{\w {\cal P}}_{\CK}
(\textbf{1}_N, \textbf{1}_N, {\bf k}_l)}{{A}(N, N, {\CM}) }\,,
\end{equation}
where ${\w {\cal P}}_{\CK}
(\textbf{1}_N, \textbf{1}_N, {\bf k}_l)$ is the number of the plane partitions (i.e., watermelon configurations) given by (\ref{cor:lenn5571}) under the limit $q\to  1$:
\begin{equation}\label{new99}
\langle \Psi(\textbf{1}_N) | {\varPi}_{\bf k} |
\Psi(\textbf{1}_N)
\rangle \,=\,
{\w {\cal P}}_{\CK}
(\textbf{1}_N, \textbf{1}_N, {\bf k}_l)\,=\,\lim_{q\to 1}
\sum_{\{{\w\bmu}_N\}_{{\bf k}_l}} S_{\w{\blad}} (\textbf{q}_N)
S_{\w{\blad}}\Bigl(
\frac{ \textbf{q}_N} {q}\Bigr)\,.
\end{equation}
The number ${ {\cal P}}_{\CK}
(\textbf{1}_N, \textbf{1}_N, {\bf 0})$ in (\ref{ratbe724}) is the number (\ref{ratbe794}) of unconstrained plane partitions in
$N\times N\times (M-N)$ box
(see Figures \ref{fig:f6} and \ref{fig:ff9}).
In the case of ${\bf k}_l= {\bf l}$, the estimate (\ref{ratbe724}) is expressed by means of ${\w {\cal P}}_{\CK}
(\textbf{1}_N, \textbf{1}_N, {\bf l})$ (\ref{cor:lin5572})
with $S_{\w {\blad}} ({\bf 1}_N)$
expressed by (\ref{numbpaths})  (cf. Section~\ref{sec53}).

The projector ${\varPi}_{\bf k}$
implies that $l$ flipped spins
of $N$ particles mean value are pinned to their positions, and thus the numbers ${\w {\cal P}}_{\CK}
(\textbf{1}_N, \textbf{1}_N, {\bf k}_l)$ enumerate the \textit{diagonally restricted} plane partitions characterized by $l$ columns of prescribed heights in the main diagonal.
The presence of the columns of fixed heights diminishes a total volume
\begin{equation}
\label{new81}
\sum_{{\bpi}\subset {\sf B}_{ N, N,{\CM}}} | {\bpi} |
\end{equation}
characterizing the set of all plane partitions
admissible for a box ${\sf B}_{N, N,{\CM}}$ of the size ${ N\times N\times {\CM}}$. The plane partitions enumerated by the numbers ${A}(N, N, {\CM} |\,n)$ (\ref{new40}) are also `diagonally restricted' since the diagonals of ${\bpi}$ subjected to $\tr_N {\bpi}=n$ also lead to a volume diminished
in comparison with (\ref{new81}).

Recall that the number of all sets of paths $|{\cal P}^0_{K, N} ({\bf k}_l)|$ (\ref{ratbe711438}) enumerates
the closed trajectories of $N$ random turns vicious walkers such that $l$ initial/final positions ${\bf k}_l$ are prescribed. In turn, the representation \eqref{new15} is the generating function of the numbers \eqref{new18} enumerating such sets of closed paths of vicious walkers that initial/final positions are labelled by partitions of certain positive integers. Both the numbers, of the lattice trajectories $|{\cal P}^0_{K, N} ({\bf k}_l)|$ (\ref{ratbe711438})
and of the diagonally restricted plane partitions ${\w {\cal P}}_{\CK}
(\textbf{1}_N, \textbf{1}_N, {\bf k}_l)$ (\ref{new99}), include summation $\sum_{\{{\w\bmu}_N\}_{{\bf k}_l}}$, which is either due to pinned initial/final positions or due to columns of fixed heights.

\section{Discussion}
\label{sec6}

The approach \cite{bmumn}, which enables to study the combinatorial implications of the quantum integrable models,
has been applied to the quantum phase model \cite{statm} and to the four-vertex model under fixed boundary conditions in the external inhomogeneous field  \cite{bmjpa}. The asymptotics of evolution of the first moment of particles distribution exponentiated has been found to
provide the norm-trace generating function of plane partitons \cite{statm}.
The partition function of the four-vertex model produced the norm-trace generating function of plane partitions \cite{st} and its generalization \cite{gan}, which describe the trace statistics of plane partitions.

The $XX$ model is of primary interest in the present paper,
and the correlation function of non-homogeneously parameterised generating exponential is studied. Generally, combinatorial implications of the $XX$ model are similar to those of the $XXZ$ chain in the limit of infinite anisotropy \cite{bmumn}. In turn, the four-vertex model is equivalent to the infinite anisotropy limit of the $XXZ$ model \cite{bmumn}.
From the viewpoint of connection
with enumerative combinatorics,
the $XX$ model as an illustrative example which enables to progress. Under various specifications the generating exponential enables obtaining of the averages of such objects as the projectors onto inconsecutive flipped spins or the powers of the first moment of flipped spins distribution.

The averages mentioned are derived in the paper in the case of long enough chain, and they are related with enumeration of the trajectories of $N$ random turns vicious walkers characterized by
restriced positions of initial/final points. The asymptotics at large value of the evolution parameter are obtained,
and the transfer occurs from enumeration of random turns walks
to enumeration of plane partitions (i.e., of watermelon configurations).

More specifically, the determinantal representation for the norm-trace generating function of boxed plane partitions with fixed height of diagonal parts is obtained
as form-factor of the generating exponential over $N$-particles states (Section~\ref{sec53}).

The transition amplitude over $N$-particles states as the generating function of $K$-step sets of random turns walks is  the main issue of a technical Section~\ref{50}. The transition amplitude is obtained in the power series form, and its coefficients fulfilling a difference equation are derived in terms of the circulant matrix expressing the $XX$ Hamiltonian.
A relationship between the entries of powers of the circulant matrix,
the lacunary sums of the binomial coefficients, and self-avoiding walks of vicious walkers is unraveled by means of the Ramus's identity and its generalizations. When the length of the chain is large enough, a connection with the problem of enumeration of increasing subsequences of random permutations is pointed out.

Two opportunities of trace definition are considered: the trace over $N$-particles Bethe states and the total trace which includes summation over numbers of particles $N$. The corresponding Boltzmann-weighted mean values are considered for the generating exponential itself, for the projector onto inconsequent flipped spins, and for a power of the first moment of flipped spins distribution.

Let us point out the new results obtained. For $N$-particles averages the estimates at large enough length of the chain are expressed through the numbers of sets of  trajectories characterized either by a subset of pinned initial/final positions or by fixed values of the whole sum of initial/final coordinates.
In the case of the total trace, the mean value of projector of inconsecutive flipped spins is presented at $M\gg 1$ as a ratio of two polynomials.
Equation \eqref{new92} demonstrates that the determinantal representation of the mean value arising at $M\to\infty$ is related with the interpretation in terms of sets of paths of random turns vicious walkers.

The $N$-particles averages are also estimated provided that the evolution parameter (inverse temperature) grows faster than the length of the chain.
The estimates are obtained in the ratio form and keep a similarity to the case of extremely long chain: although the sets of random turns trajectories are replaced, at large evolution parameter, by plane partitions, the restrictions imposed look similar. The nominators are given by the numbers of the diagonally restricted plane partitions which are either in one-to-one correspondence with the flipped spins positions or characterized by fixed trace of all diagonal elements. The denominators correspond to generic plane partitions.

The results obtained look stimulating from the viewpoint
of further investigation of the four-vertex model, of the phase model, and of the $XY$ model (cf.~\cite{gauss}) along the lines presented.

\section*{Acknowledgement}

Supported by RSF (No.~18-11-00297).

\section*{Appendix I}

{\sf Proposition~3} is devoted to the verification of the representation (\ref{qanal272}) expressing the number of sets of paths of random turns vicious walkers. The corresponding difference equation
(\ref{mpcf6}) is a tool of verification rather than derivation of (\ref{qanal272}), as stressed in \cite{forr2}.
The present {\sf Appendix~I}
is concerned with the derivation by means of (\ref{qanal13}).

Let us begin with the derivation of the relation
\begin{equation}
\mathfrak{G}^{\,0}({\bmu^L}; {\bmu^R} | K) = \sum_{|{\bf n}|=K} P({\bf n})\,
{\bold\Delta}^{\bf n}_{{\bmu^L}; {\bmu^R}}\,,
\nonumber
\eqno({\rm AI.1})
\end{equation}
where ${\bf n}=(n_1, n_2, \ldots, n_N)$, $|{\bf n}|\equiv n_1+n_2+ \ldots + n_N$, $P({\bf n})$ is the multinomial coefficient  (\ref{qanal271}), and ${\bold\Delta}^{\bf n}_{{\bmu^L}; {\bmu^R}}$ is given by (\ref{qanal27}). The commutation relation (\ref{comcom}) supplied with $H_{\rm xx} \mid \Uparrow\rangle = 0$ and $\si^z_k \mid \Uparrow\rangle =\mid \Uparrow\rangle$
enables us to obtain (AI.1) at $K=1$ as the base case of induction. As induction step, it is assumed that (AI.1) is valid at $K-1$. We put $H_{\rm xx}^{K}= H_{\rm xx}^{K-1} H_{\rm xx}$ in (\ref{qanal13}) to prove (AI.1) and obtain:
$$
\begin{array}{rcl}
\displaystyle{
\mathfrak{G}^{\,0}({\bmu^L}; {\bmu^R} | K)} &=& \displaystyle{
\sum_{l=1}^{N}\sum_{k=1}^M
{\bold\Delta}_{\mu^R_l,\,k}
\sum_{|{\bf n}|=K-1} P({\bf n})\,
{\bold\Delta}^{\bf n}_{{\bmu^L}; {(\mu^R_1, \ldots, \mu^R_{l-1}, k, \mu^R_{l+1}, \ldots, \mu^R_N) }}}
\\
&=&\displaystyle{
\sum_{l=1}^{N} \sum_{|{\bf n}|=K-1} P({\bf n})\, {\bold\Delta}^{{\bf n}+{\bf e}_{l}}_{{{\bmu}^L}; {{\bmu^R}}}
}\,,
\end{array}
\eqno({\rm AI.2})
$$
where $N$-tuples ${\bf e}_{l}$ are defined in (\ref{ratbe7272}).
The multinomial theorem demonstrates that (AI.2) leads to (AI.1). The determinantal generalization of (AI.1) leads to
(\ref{qanal272}), where the non-intersection requirements
are taken into account.

\section*{Appendix II}

The Ramus's identity \cite{ram1} is of the form:
\begin{equation}
\nonumber
\frac{2^{n}}{R}\sum_{j=0}
^{R-1}
\cos^n\frac{\pi j}{R}
\cos\frac{\pi j (n-2t)}{R}
\,=\,
\begin{pmatrix} n \\ t \end{pmatrix} +
\begin{pmatrix} n \\ t + R \cdot 1 \end{pmatrix} +
\begin{pmatrix} n \\ t + R \cdot 2 \end{pmatrix} + \ldots \,,
\eqno({\rm AII.1})
\end{equation}
where $0\le t < R$.

\section*{Appendix III}

It is straightforward to obtain
useful identities provided that the expressions for $\bigl({\bold\Delta}^K \bigr)_{j m}$ given by \textsf{Proposition~4}, on one hand,
and by \cite{rim1, rim2}, on another, are equated each to other. Without reproducing the appropriate formulae from \cite{rim1, rim2}, we simply specify, according to \textsf{Proposition~4}, the matrix $\bigl({\bold\Delta}^K \bigr)_{j m}\equiv \bigl({\bold\Delta}^K \bigr)_{j - m}$ of the size $6\times 6$ to $K=14$:
\begin{equation}
\begin{array}{l}
\bigl({\bold\Delta}^{14} \bigr)_{0} = \begin{pmatrix} 14 \\ 1 \end{pmatrix}_{3} = 5462\,,\\ [0.3cm] \bigl({\bold\Delta}^{14} \bigr)_{2} = \begin{pmatrix} 14 \\ 0 \end{pmatrix}_{3} =  \bigl({\bold\Delta}^{14} \bigr)_{4} = \begin{pmatrix} 14 \\ 2 \end{pmatrix}_{3} = 5461\,,\\[0.3cm]
\bigl({\bold\Delta}^{14} \bigr)_{1} =
\bigl({\bold\Delta}^{14} \bigr)_{3} = \bigl({\bold\Delta}^{14} \bigr)_{5} = 0 \,.
\end{array}
\nonumber
\end{equation}
We obtain in notations \cite{rim1, rim2}:
\begin{equation}
\begin{array}{l}
\bigl({\bold\Delta}^{14} \bigr)_{0} =
\displaystyle{\frac{a_1}{6} = \frac{2(2^{14}+2)}{6}}\,,\\[0.3cm]
\bigl({\bold\Delta}^{14} \bigr)_{2} = \bigl({\bold\Delta}^{14} \bigr)_{4} = \displaystyle{ \frac{a_3}{6}=
 \frac{2(2^{14}-1)}{6}
 = \frac{a_1}{6} - 1} \,.
\end{array}
\nonumber
\end{equation}

Further, we specify  $\bigl({\bold\Delta}^K \bigr)_{j - m}$ to $M=6$ and $K=15$:
\begin{equation}
\begin{array}{l}
\bigl({\bold\Delta}^{15} \bigr)_{1} = \begin{pmatrix} 15 \\ 1 \end{pmatrix}_{3} =  \bigl({\bold\Delta}^{15} \bigr)_{5} = \begin{pmatrix} 15 \\ 2 \end{pmatrix}_{3} = 10923
\,,\\ [0.3cm]
\bigl({\bold\Delta}^{15} \bigr)_{3} = \begin{pmatrix} 15 \\ 0 \end{pmatrix}_{3} = 10922
\,,\\[0.3cm]
\bigl({\bold\Delta}^{15} \bigr)_{0} =
\bigl({\bold\Delta}^{15} \bigr)_{2} = \bigl({\bold\Delta}^{15} \bigr)_{4} = 0 \,.
\end{array}
\nonumber
\end{equation}
As well,
\[
\bigl({\bold\Delta}^{15} \bigr)_{1}
 = \bigl({\bold\Delta}^{14}
\bigr)_{0} + \bigl({\bold\Delta}^{14} \bigr)_{2}\,, \qquad \bigl({\bold\Delta}^{15} \bigr)_{3}
= \bigl({\bold\Delta}^{14} \bigr)_{2} + \bigl({\bold\Delta}^{14} \bigr)_{4}\,.\]
We obtain in notations \cite{rim1, rim2}:
\begin{equation}
\begin{array}{l}
\bigl({\bold\Delta}^{15} \bigr)_{1} = \bigl({\bold\Delta}^{15} \bigr)_{5} = \displaystyle{ \frac{a_2}{6}=
\frac{2(2^{15}+1)}{6}}\,,\\[0.3cm]
\bigl({\bold\Delta}^{15} \bigr)_{3} =
\displaystyle{\frac{a_4}{6} = \frac{2(2^{15}-2)}{6}
 = \frac{a_2}{6} - 1} \,.
\end{array}
\nonumber
\end{equation}

\end{document}